\documentclass[a4paper,fleqn]{cas-dc}

\usepackage{xurl}
\usepackage{float}
\usepackage[round]{natbib}
\usepackage{graphicx}
\usepackage{placeins} 
\usepackage[nottoc,numbib]{tocbibind} 
\usepackage{doi}

\usepackage{pdfpages}

\usepackage{chngcntr}    

\usepackage{arydshln}
\usepackage{gensymb} 

\AtBeginDocument{%
 \hypersetup{
  citecolor=Bittersweet,
    linkcolor=Blue,
  urlcolor=Blue}
}

\makeatletter
\def\ps@first{%
   \let\@oddhead\@empty
   \let\@evenhead\@empty
   \def\@oddfoot{}
   \let\@evenfoot\@oddfoot
}
\ExplSyntaxOn
\cs_gset:Npn \__first_footerline:
  { \group_begin: \small \sffamily \__short_authors: \group_end: }
\ExplSyntaxOff

\begin{document}

\title[mode = title]{Transition dynamics of electricity asset-owning firms}
\shorttitle{Transition dynamics of electricity asset-owning firms}

\author[1,2]{Anton Pichler}[orcid=0000-0002-7522-1532]
\cormark[1]
\cortext[cor1]{Corresponding author: anton.pichler@wu.ac.at}

\address[1]{Vienna University of Economics and Business, Vienna, Austria}
\address[2]{Macrocosm Inc, New York, USA}

\shortauthors{Pichler}

\date{\today}

\begin{abstract}
\noindent Despite dramatic growth and cost improvements in renewables, existing energy companies exhibit significant inertia in adapting to the evolving technological landscape. This study examines technology transition patterns by analyzing over 140,000 investments in power assets over more than two decades, focusing on how firms expand existing technology holdings and adopt new technologies.
Building on our comprehensive micro-level dataset, we provide a number of quantitative metrics on global investment dynamism and the evolution of technology portfolios. We find that only about 10\% of firms experience capacity changes in a given year, and that technology portfolios of firms are highly concentrated and persistent in time. We also identify a small subset of frequently investing firms that tend to be large and are key drivers of global technology-specific capacity expansion. 
Technology transitions within companies are extremely rare. 
Less than 3\% of the more than 8,400 fossil fuel dominated firms have substantially transformed their portfolios to a renewable focus and firms fully transitioning to renewables are, up-to-date, virtually non-existent. 
Notably, firms divesting into renewables do not exhibit very characteristic technology-transition patterns but rather follow idiosyncratic transition pathways. 
Our results quantify the complex technology diffusion dynamics and the diverse corporate responses to a changing technology landscape, highlighting the challenge of designing general policies aimed at fostering technological transitions at the level of firms.
\end{abstract}


\maketitle

\section{Introduction} \label{sec:intro}

As decarbonization pressures remain high, the technology landscape in the electricity sector is poised for transformation.
Typically, energy transition scenarios are investigated at the level of technologies that neglect the role of individual decision-making agents, such as firms. However, incumbent stakeholders face substantial transition risks with potential adverse impacts beyond their own operations \citep{caldecott2015stranded,battiston2017climate, mercure2018macroeconomic, hickey2021can}. Moreover, incumbent firms could use their market power to delay the transition \citep{fouquet2016historical}.
Consequently, a better understanding of how firms drive or resist the energy transition is crucial for investors, financial regulators, policymakers, and the affected firms themselves.

Independent power producers (IPPs) and other firms hold substantially larger shares of renewable energy than electric utilities, which have traditionally dominated electricity markets and continue to represent the largest owners of fossil power assets (Supplementary Information [SI] Fig.~S4).
However, staying within the carbon budgets necessary for avoiding extreme climate change will involve the stranding of substantial amounts of fossil assets \citep{davis2010future, mcglade2015geographical, pfeiffer20162, pfeiffer2018committed, tong2019committed, lu2022plant, rekker2023evaluating}. Moreover, the heterogeneous exposures to renewable and fossil assets \citep{baer2022trisk} and the high levels of ownership concentration \citep{semieniuk2022stranded, von2023concentration} will likely result in diverse transition impacts across different regions.

A fast-changing technology landscape can threaten the existence of incumbent firms if they fail to adapt their value propositions and business models \citep{christensen2015innovator, richter2012utilities, castaneda2017evaluating}. 
To gain a comprehensive understanding of the socio-economic impacts of the energy transition, it is crucial to examine how fossil-focused energy companies can adjust and innovate their business models.
Decarbonizing fossil-based portfolios through investments in low-carbon energy is an apparent adaptation strategy in the transitioning energy sector \citep{richter2013business, bryant2018typologies}. However, while major utilities are investing in renewables \citep{patala2021multinational,steffen2022state}, case studies suggest that these efforts fall short of transitioning towards low-carbon technology portfolios \citep{kungl2015stewards}, suggesting that the energy transition follows a ``substitution'' path where challenger firms crowd out incumbent utilities \citep{kelsey2018wins}.

Despite the relevance of this topic, large-scale firm-level studies on technology adoption in the power sector are scarce. A notable exception is \cite{alova2020global}, which examines the technology portfolio dynamics of more than 3k electric utilities. The study shows that while many utilities have adopted renewable energy technologies, the majority continue to expand or maintain their fossil fuel assets.

In this paper, we contribute to this literature by providing a comprehensive quantitative characterization of the diffusion of key energy technologies at the level of firms, focusing on empirical transition and diversification pathways of previously fossil-dominated companies. Using an extensive asset-level database covering more than 70k power plants and 20k firms (see Appendix~\ref{apx:data}), we show that energy assets are highly unevenly distributed across firms and that firms' technology portfolios are strongly biased towards single technologies. 
Technology portfolios are highly static in time, as only a small minority of firms invest in new assets in any given year. We can identify a small subgroup of, typically, large firms that are investing very frequently. These frequent investors account for a large portion of the observed expansion of electricity generation capacity. However, even if firms invest, their technological choices resonate with existing portfolio concentrations, resulting in a highly path-dependent evolution of technology portfolios.

Despite persistent investment dynamics, we show that fossil-focused businesses have contributed non-negligibly to the recent expansion of renewable energy. Nevertheless, only a small minority of them have significantly transformed their portfolios from a fossil to a renewable focus. When zooming into the investment dynamics of this small subset, we do not find highly typical technology transition pathways. Instead, we find diverse diversification and transition strategies of firms. While the adoption of solar and wind power represents the most crucial component of technology focus-switching companies, we find that the adoption of hydro- and biopower also played an important role.

\section{Results} \label{sec:results}

\begin{figure*}[!htb]
    \centering
    \includegraphics[width = \textwidth]{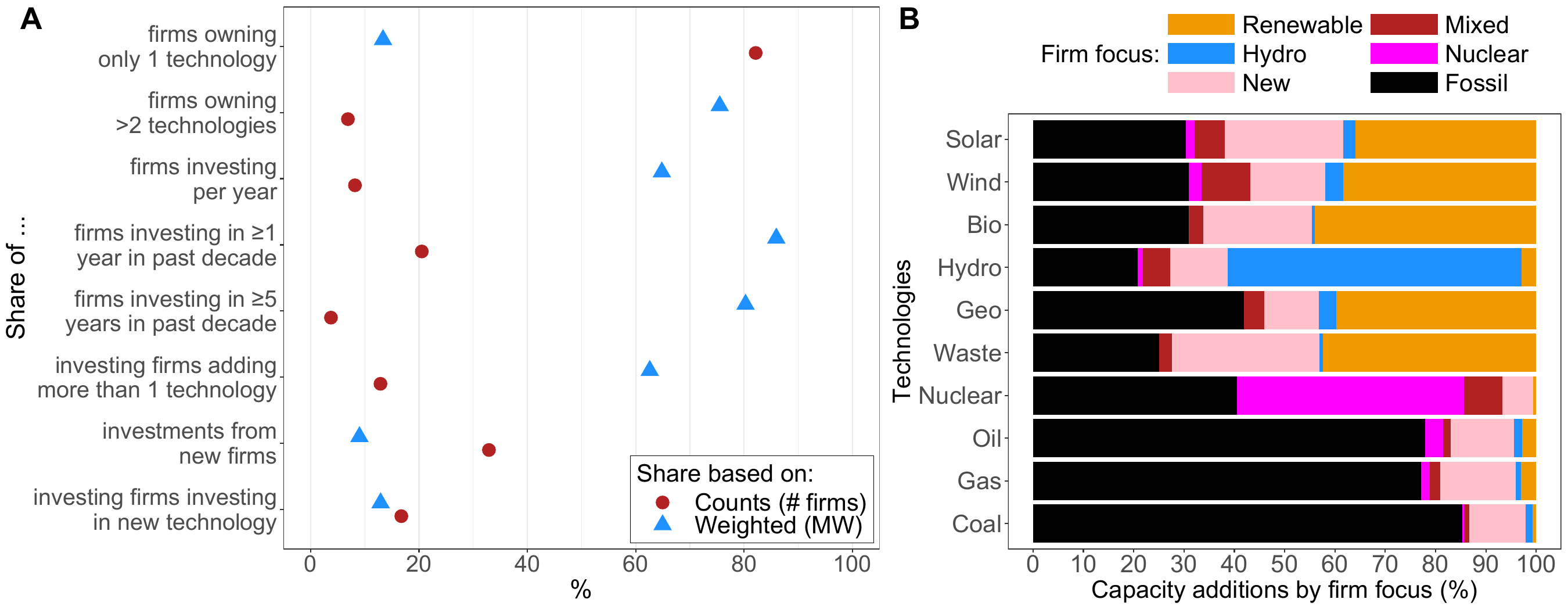} 
    \caption{
    {\bf A. Technology concentration and investment dynamics.}
The dots represent the average shares per year computed over the 2001--2022 horizon. Line ranges indicate the minimum and maximum shares per year over the same horizon.
Shares computed based on counts (either the number of firms or a number of investments) are shown in circles and with red color. Blue-colored triangles represent shares computed based on capacity owned or capacity invested.
 {\bf B. Technology expansion by firm focus over the past decade (2013--2022).} 
 Each row on the y-axis represents a specific technology. The x-axis shows the share of capacity expansion due to investments by the different firm focus types (see colors). \emph{New firms} are firms that have not held any power assets prior to 2013. \emph{Renewable firms} are firms which portfolios consists of at least 50\% renewables. \emph{Nuclear} and \emph{fossil firms} are defined analogously. \emph{Mixed firms} do not hold more than 50\% of renewables, fossils, or nuclear power. 
    }
    \label{fig:result1}
\end{figure*}

\subsection{Uneven distribution of power assets}
\label{sec:owner}

We focus on asset ownership at the level of ultimate parent companies, as direct owners frequently only act as their operating subsidiaries. An analysis at the level of direct owners thus might mask the diversification strategies of companies through specialized subsidiaries (Appendix~\ref{apx:owner_map}). We further do not restrict our sample to utilities or IPPs, as is done in previous studies, as we observe that other company types hold more than 30\% of total capacity (SI~Section~S3.1).

While we observe a large number of power asset-owning legal entities ($\sim$23k ultimate parents and $\sim$30k direct owners), the bulk of these assets is concentrated in a few hands. 
80\% of all installed power capacity is held by less than 3\% of firms in our sample, corresponding to merely 633 firms (SI~Section~S3.2). 
We find that the top 25 companies own about a quarter of the total installed capacity. 
Compared to the global average, the top 25 firms own a disproportionately large share of coal power, accounting for about 40\% of worldwide installed coal power capacity.

The development of solar and wind power is often attributed to small-scale firms. However, we observe similar ownership distributions across all major technology groups -- all of them show concentration in large firms, and renewables are not an exception. Moreover, the ownership of power capacities has not become less evenly distributed over time as the increasing penetration of renewables might have suggested. 
Thus, the global ownership of power assets is dominated by a relatively small number of key players who are likely to play an important role in the energy transition.

\subsection{High technological concentration}
\label{sec:techfocus}
If we zoom into the technological composition of the firms' assets, we find extremely high levels of concentration. About 80\% of firms hold only a single technology, so multi-technology energy firms are the exception rather than the rule (Fig.~\ref{fig:result1}A). However, a clear positive relationship exists between firm size and the number of technologies owned (SI~Section~S3.3). The 20\% of firms owning multiple technologies own almost 90\% of the total capacity in our sample. While the typical single technology firm is small, and the typical multiple-capacity firm is large, there are also notable exceptions. Among them are the Venezuelan Electrificaci\'on del Caron\'i or the Iranian Water and Power Resources Development Company, which hold multiple GW of hydropower-only. We observe similar notable exceptions across various technological domains. For example, the typical solar PV-dominated firm is comparatively small (SI~Section~S3.3). Nevertheless, in our sample, there are eleven solar PV-only firms larger than 1~GW (e.g., Canadian Solar Inc., Pine Gate Renewables). 

Interestingly, even multiple-technology firms tend to be highly specialized in single technologies. 
The typical 2-technology firm allocates almost 80\% of its total capacity to a primary technology rather than evenly splitting it between both technologies.
Similarly, the dominant technology of the median 3-technology firm accounts for about 70\% of portfolio capacity, compared to an even 33\%-split across the three technologies. We observe similar high levels of technology concentrations for firms owning even more technologies (SI~Section~S3.4).

\subsection{Slow and persistent investment dynamics}
\label{sec:invest}

Confirming previous research \citep{alova2020global}, we find that electricity technology portfolios change slowly over time.
In an average year, less than 10\% of firms in our sample invest in new capacity, and over the past decade, only about a fifth of firms have made any investments (Fig.~\ref{fig:result1}A). However, the inclination to invest is strongly size-dependent: The firms that do invest represent about two-thirds of global installed capacity. Investments are primarily focused on single technologies. Only a minority of (typically large) firms invest in more than one technology per year (Fig.~\ref{fig:result1}A).

We run logistic regressions to investigate which firm characteristics affect the probability of investing (SI~Section~S3.5). While firm size plays a role, the two most important covariates are a firm's historical investment frequency and the number of technologies it owns. We identify a small fraction of persistent investors who make capacity additions on a regular basis. About 4\% of firms have invested in at least half of the years over the past decade. These companies account for $\approx 80\%$ of global power capacity (Fig.~\ref{fig:result1}A).  
A firm's technology focus also affects its inclination to invest: Firms focusing on solar PV and wind onshore are more likely to invest than fossil firms or firms with more diversified portfolios.

Recent market entrants such as IPPs play a crucial role in the deployment of renewable energy and challenge the dominance of incumbent utilities \citep{kelsey2018wins}. We identify high entry rates into the power sector, with one-third of all investing firms corresponding to new market entrants (i.e., firms that did not hold power assets previously). These new firms account for about 10\% of all new capacity additions in any given year (Fig.~\ref{fig:result1}A).
As demonstrated in Fig.~\ref{fig:result1}B, firms entering the market over the last decade account for about 30\% of total expansion in solar, biomass, and waste power plants.

A crucial question in understanding energy transition dynamics is whether firms invest in new technologies, i.e., technologies that they have not owned previously. About 15-20\% of investing firms add new technologies in a given year. Taken together with the low investment rates, this implies that the rate of new technologies entering existing technology portfolios is small: If 10\% of firms invest in a given year and 15\% of them invest in new technologies, only 1.5\% of existing portfolios experience additions of new technologies in any year. 

In SI~Section~S3.6, we fit logistic regression models to investigate potential driving factors behind the adoption of specific technologies. 
Our results show that a technology's relative share within a firm's portfolio is a key factor in determining whether an investor adds capacity of that particular technology. 
This relationship is nonlinear, becoming stronger with higher portfolio concentration.
The second key explanatory variable is the historical investment behavior. The likelihood of investing in a particular technology increases substantially if an investment in that technology has been made in the preceding years, particularly if a firm invested exclusively in that technology. 
These simple regression specifications explain a substantial share of the observed variance, indicating that an investor's technology choice is relatively predictive. Specifically, investors' technology choices mimic previous choices and strongly reinforce existing portfolio compositions. 

Despite strong path dependencies in firms' investment choices, fossil-dominated companies have substantially contributed to the expansion of low-carbon technologies over the past decade (Fig.~\ref{fig:result1}B). Over the past ten years, about 40\% of all added solar, wind, and biomass capacity is due to fossil companies, and their contribution to geothermal expansion is even larger. Conversely, we observe that investments in fossil technologies are much more concentrated and dominated by fossil firms. Renewable firms have contributed only little to new developments of fossil capacities and not at all to nuclear power expansion. Note that this is also due to size effects. Fossil firms tend to be larger than renewable firms (SI~Fig.~S6A) and investment amount strongly correlates with firm size (SI~Section S3.6).

\begin{figure*}[!htb]
    \centering
    \includegraphics[width = \textwidth]{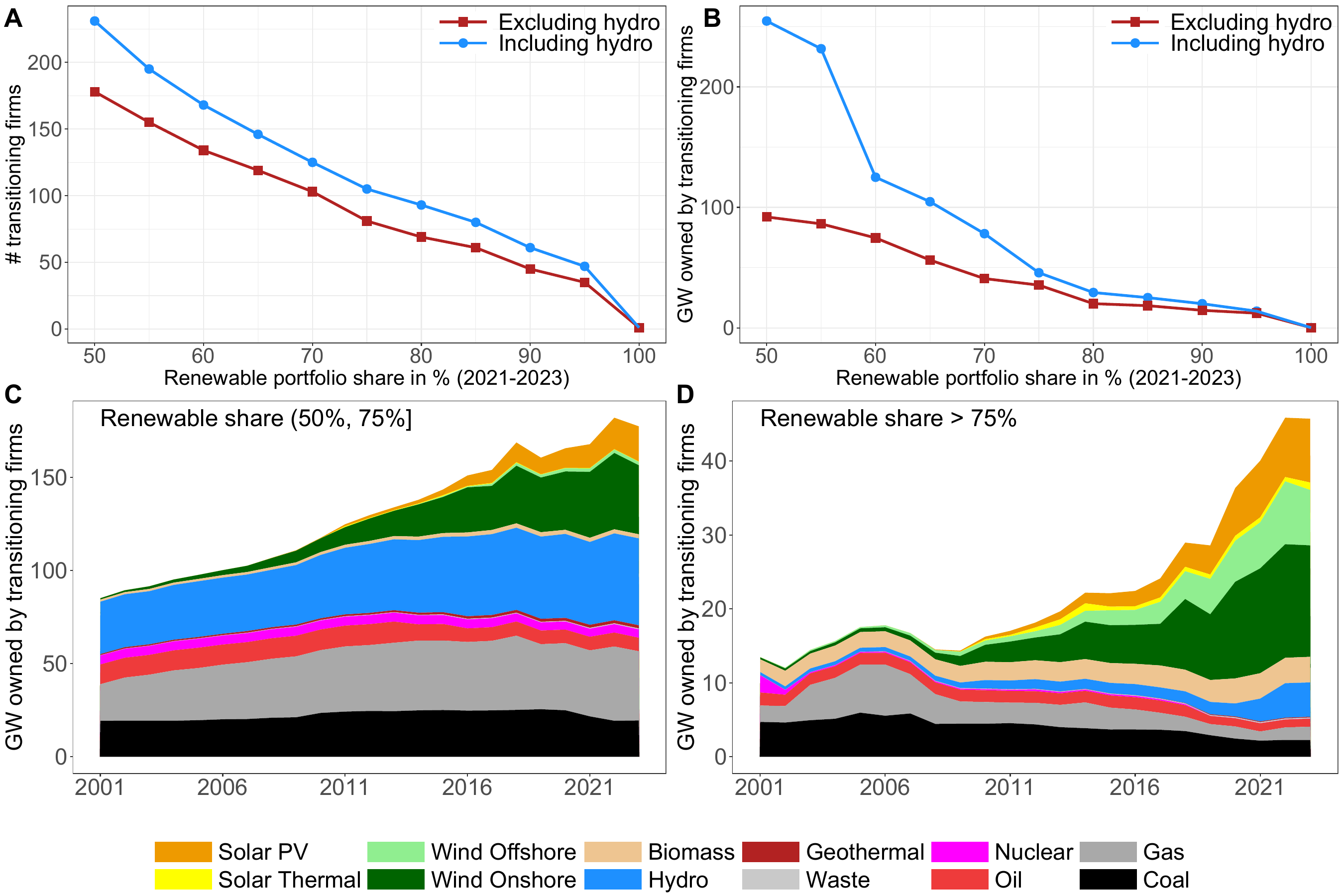} 
    \caption{
    {\bf A. Number of transitioning firms for different renewable portfolio share thresholds.}
        {\bf B. Total capacity (in GW) owned by transitioning firms in 2023 for different renewable portfolio share thresholds.}  
    To qualify as transitioning, a firm must have had at least 50\% of its portfolio in fossil-based capacity for a minimum of five consecutive years.
    The x-axis represents the threshold indicating whether a firm held at least x-\% of renewable capacity in its portfolio for every year between 2021 and 2023.  
    The blue (red) line includes (excludes) hydropower in the definition of renewables.
    {\bf C and D. Technology-specific capacities of transitioning firms over time.}
    In C, the sample is restricted to transitioning firms that held between 50\% and 75\% of renewables between 2021 and 2023. In D, the sample is restricted to firms holding more than 75\% renewables during the same period.
    }
    \label{fig:result2}
        \end{figure*}

\begin{figure*}[!htb]
    \centering
    \includegraphics[width = \textwidth]{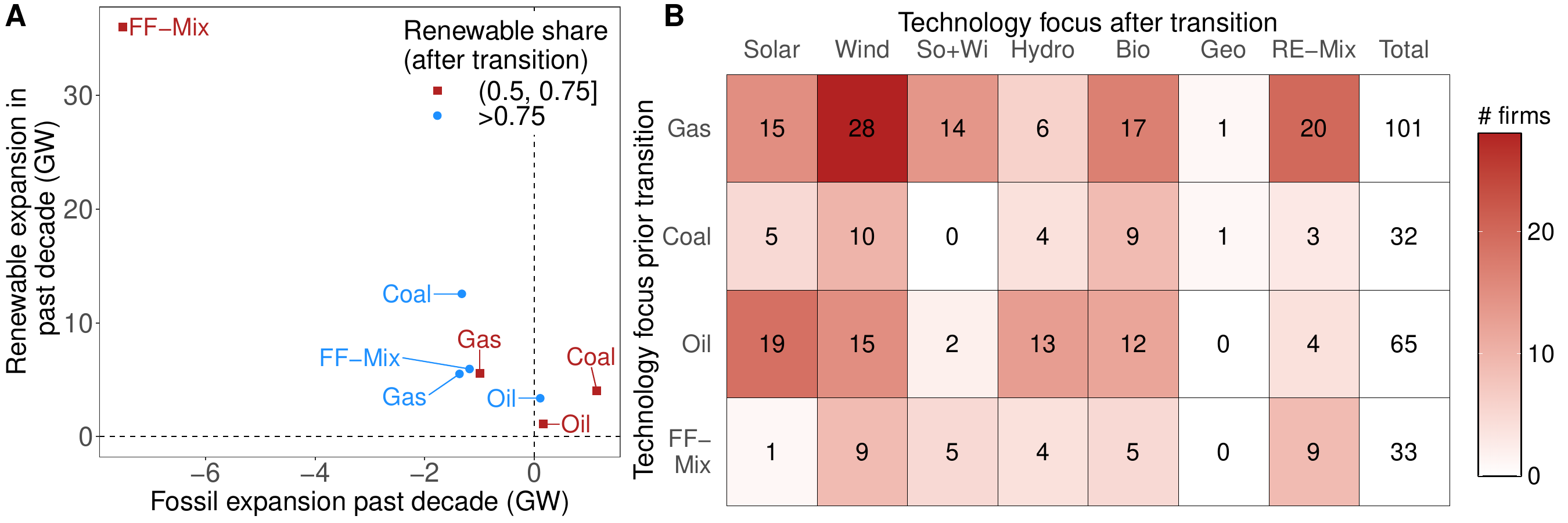} 
    \caption{
    {\bf A. Renewable and fossil expansion over past ten years of transitioning firms, based on their technology focus prior to transitioning.} 
    A technology focus of a firm is defined as the technology accounting for more than 50\% of its technology portfolio. 
Colors distinguish between firms with 50–75\% renewables in their portfolio and those with over 75\%, after transitioning between 2021 and 2023.
    {\bf B. Technology transition pathways.}
    The y-axis indicates firms' technology focus prior to transitioning, and the x-axis indicates their technology focus after transitioning (using the 50\% renewable share threshold).
    Numbers and colors represent the number of firms transitioning (\emph{Total} is not colored).
    \emph{So+Wi} are firms that do not hold a majority in a single technology class but a majority in wind and solar capacity combined.
    In both figures, \emph{FF-Mix (RE-Mix)} indicates firms that hold more than 50\% of fossil (renewable) capacity, but no single technology (neither solar+wind) dominates.
    } 
    \label{fig:result3}
        \end{figure*}

\begin{figure*}[!htb]
    \centering
    \includegraphics[width = \textwidth]{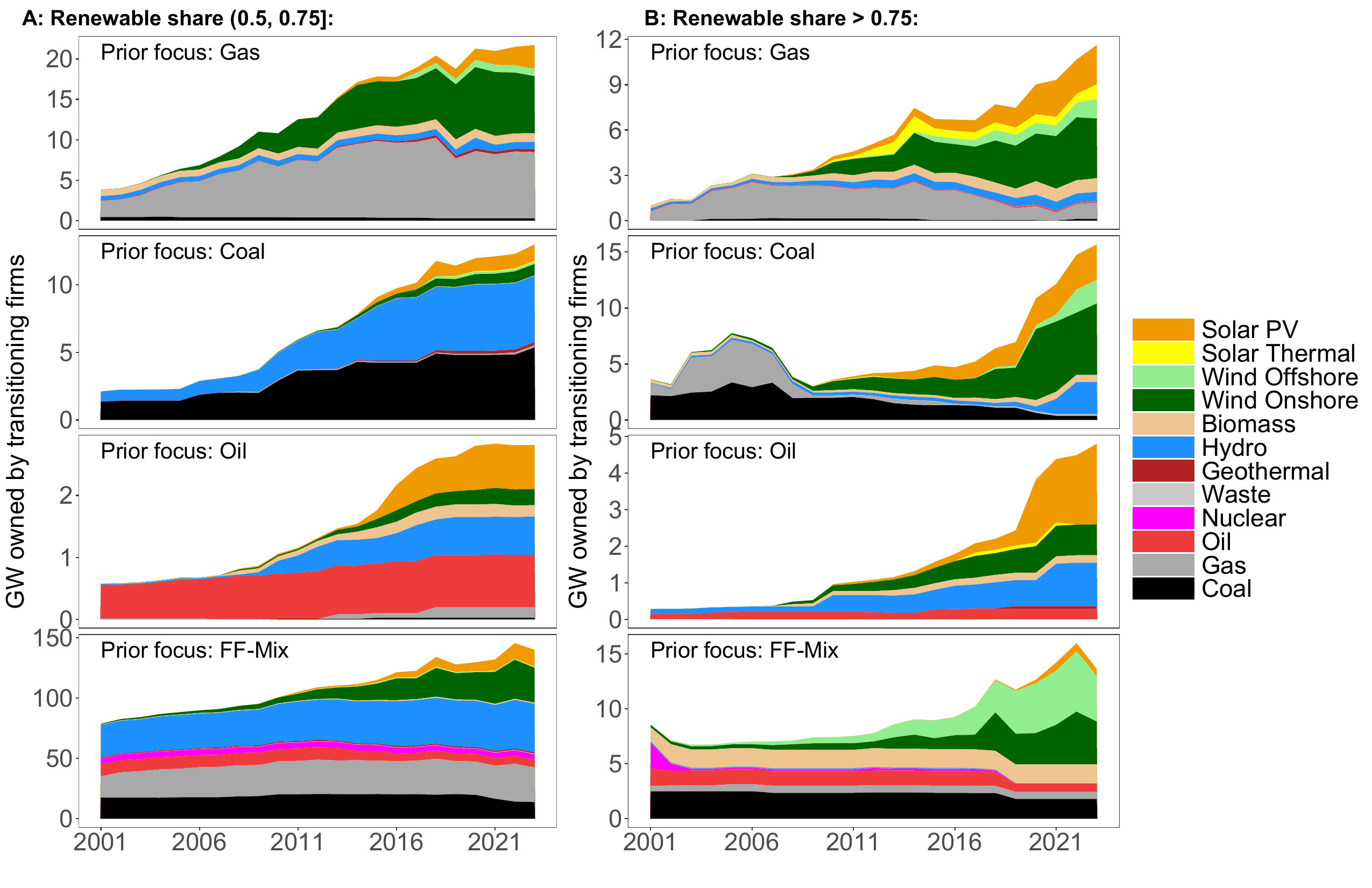} 
    \caption{
    {\bf Capacity evolution of transitioning firms with different prior technology focus.} 
A. The sample is restricted to transitioning firms that held between 50\% and 75\% of renewables between 2021 and 2023 (moderate transition). B. The sample is restricted to firms holding at least 75\% renewables (radical transition). \emph{Prior focus} of a firm corresponds to its focal fossil technology accounting for >50\% of total capacity prior to the transition. Diversified fossil companies without a single focal technology are denoted as \emph{FF-Mix}.
    } 
\label{fig:result4}
\end{figure*}

\subsection{Transition dynamics of firms} \label{sec:firm_trans}

We consider a firm having transitioned from fossils to renewables if the firm's total capacity consisted of $\ge$50\% fossil power (coal + gas + oil) for five consecutive years in the past but held a majority in renewables (solar, wind, hydro, geothermal, biomass) in every year between 2021 and 2023 (see Appendix~\ref{apx:trans} for details). 

Following this definition, Fig.~\ref{fig:result2}A shows the number of transitioning firms for different renewable portfolio share thresholds. Out of more than 8,400 companies that have held predominantly fossils for at least five years, we can identify about 230 firms (2.7\%) that have transitioned to at least 50\% renewables. These firms held about 160 GW when exceeding the 50\% renewable threshold and grew to 255 GW in 2023 (Fig.~\ref{fig:result2}B), accounting for 3.5\% of global capacity.

If we focus only on non-hydro renewables, the number of fossil companies transitioning drops to only 178, indicating that the majority of transitioning firms do not rely on adopting hydropower. However, transitioning firms that rely on adopting hydropower tend to be very large (Fig.~\ref{fig:result2}B). 
A prominent example of a large transitioning firm reliant on hydropower is the Italian multinational utility, Enel. Over the past decade, Enel has shrunk its fossil portfolio share from almost 60\% to 35\% by expanding its capacity in hydro, solar PV, and onshore wind while at the same time decreasing its coal and gas power fleets (after growing them previously). In Nov 2024, Enel announced to invest an additional EUR~12bn into renewable energy expansion until 2027 \citep{enel}.

Figs.~\ref{fig:result2}A-B make clear that the number of transitioning firms declines if we apply more stringent definitions. If we require previously fossil-based companies to increase their renewable shares to at least 75\%, the number of transitioning firms drops to less than half, owning less than 50 GW of power capacity in 2023. Strikingly, despite the long time horizon and large sample, only one previously fossil-fuel-dominated firm has transitioned to 100\% renewables: the Finnish company Oy Turku Energia. This transition was achieved by retrofitting the previously coal-powered Naantali CHP plant to use biomass.

After transitioning, firms that hold between 50\% and 75\% renewables own large amounts of hydropower and have also vastly expanded their onshore wind capacities. However, they have not unwound their fossil assets in absolute terms (Fig.~\ref{fig:result2}C).
If we restrict the sample to firms that have switched to >75\% renewables, the important role of solar PV and offshore wind capacity becomes visible. In contrast, hydro capacities play a limited role for these firms (Fig.~\ref{fig:result2}D). Importantly, these firms have significantly reduced their fossil assets to almost a third over the past ten years.

Our results indicate that few firms have transitioned from fossil fuel technologies to renewables. While a number of previously fossil-dominated companies have expanded their renewable capacities, most still maintain significant fossil fuel assets. These firms have diversified into renewables rather than fully transitioning away from fossil fuels.

\subsection{Technology transition dynamics} \label{sec:tech_trans}

Among transitioning fossil firms, previously gas-focused companies represent the largest group (Fig.~\ref{fig:result3}B).
Conversely, diversified fossil firms, i.e., firms that do not hold an absolute majority of a single fossil technology, represent a relatively small group. Only 33 of them meet our criteria for transitioning to renewable energy. However, these firms tend to be large and have significantly contributed to the expansion of renewable capacity at the expense of fossil capacity (Fig.~\ref{fig:result3}A). Transitioning diversified fossil companies -- that have only moderately transformed their portfolios to renewable shares between 50\% and 75\% -- have expanded more than 35 GW of renewables and divested from more than 7 GW of fossil power capacity over the past decade.

Transitioning oil-focused firms have continued to expand their fossil capacity over the past decade. Similarly, transitioning coal-focused firms, which now hold between 50\% and 75\% renewables, have increased their fossil capacity. In contrast, all other transitioning firm types have reduced their fossil capacity ownership in absolute terms while simultaneously expanding their renewable energy assets.

Fig.~\ref{fig:result3}B makes it clear that there are no ``typical'' transition pathways. Instead, there is substantial heterogeneity in how transitioning firms move in the technology space. Most transitioned portfolios are dominated by solar and wind power. Despite accounting only for a small share of global installed capacity, we also observe relatively many transitions to biopower. This observation may be explained by the fact that biopower is combustion-based, making it easier for firms specialized in fossil generation to adopt.

Previously mixed fossil companies tend to transition to diversified renewable portfolios or focus on wind power. A well-known example of this group is \O{}rsted A/S, which, based on our typology, was a diversified fossil company before transitioning to a diversified renewable focus in the early 2010s. Through continued expansion of (off- and onshore) wind power, \O{}rsted now holds a majority of wind power in its energy portfolio.

Firms that have transitioned to renewables only \emph{moderately} have dramatically increased their solar PV and wind power ownership (Fig.~\ref{fig:result4}A). 
Particularly in transitioning gas companies, the expansion of solar and wind energy dwarfs the increase in other renewable capacities. Moreover, coal- and oil-based companies have substantially increased their stakes in hydropower. While hydropower also plays a significant role for previously diversified fossil companies (FF-Mix), they have predominantly grown wind and solar power. Notably, these firms still hold on to substantial amounts of fossil capacity. 

Firms that have transitioned to renewables more \emph{radically} have started unwinding their fossil capacities (Fig.~\ref{fig:result4}B). Still, these firms have grown in absolute terms due to the massive expansion of solar and wind power. For these firms, the expansion of hydropower only played a minor role in the transition. 

Notably, diversified fossil firms that have only moderately transitioned represent by far the largest group in terms of capacity ownership. Although they still hold a minority share of wind and solar capacity in recent years, their expansion in these technologies has been enormous in absolute terms, surpassing the growth achieved by other types of transitioning firms.

\section{Discussion} \label{sec:discuss}
This paper contributes to the recent research on firm-level transition dynamics in the energy sector \citep{bryant2018typologies,kelsey2018wins, alova2020global, alova2021machine, steffen2022state}. 
Our results demonstrate that firm power technology portfolios are highly concentrated, with most firms focusing on single technologies.
Firm investment dynamics are sluggish, as only a small share of power asset-owning firms invest in a given year. However, there exists a minority of frequent investors, typically larger firms, that are crucial in driving the expansion of global power capacity.
In line with institutional inertia and technological lock-in \citep{nelson1982evolutionary, arthur1989competing, unruh2000understanding}, we find that investment decisions exhibit strong path dependency, with existing portfolio concentrations and previous technology choices being the key factors in explaining future technology adoption.

The dramatic cost declines in renewable energy over the past two decades \citep{way2022empirically} have not led to systematic transformations of fossil-focused technology portfolios. While major incumbent firms have started integrating renewables into their technology portfolios \citep{frei2018leaders, alova2020global} and have contributed substantially to the expansion of renewables, they still tend to hold on to fossil-dominated portfolios that are non-compliant with 1.5\degree C  pathways \citep{rekker2022measuring}.

We have also identified several exceptions where fossil fuel firms shifted their focus toward renewable technologies. While the expansion of wind and solar capacity played a significant role for transitioning firms, we did not observe very typical transition trajectories tied to firms' existing technological predispositions. Transition pathways appear idiosyncratic, depending on the specific historical context and business case. 
This poses challenges for designing policies targeted at fostering firm transitions to renewable energy. Therefore, conducting more detailed, in-depth analyses of individual transition trajectories will be necessary to better understand potential drivers and barriers influencing these transitions.

Limitations of this study point to interesting future research avenues. For example, we did not include financial characteristics of firms or policy constraints in our analysis, as this would have compromised the coverage of our data. Clearly, financial aspects such as rates of return, financing costs, and debt structures affect investment decisions and the adaptive capacity of firms in the energy sector \citep{polzin2015public, egli2018dynamic, egli2020renewable, zhou2021energy, polzin2019policies, kempa2021cost} and could enrich our technology-centered analysis.

Further work could also be directed towards building predictive firm-level transition models. Our analysis reveals that technology choices and investment dynamics are relatively predictive, indicating the potential to forecast the future evolution of firms' technology portfolios. Such predictions could complement existing business-as-usual scenarios obtained from top-down projections and yield more granular insights into asset stranding and firm-level transition risks.

\section*{Acknowledgments}
This work was supported by the OeNB anniversary fund under contract number 18943 and the NSF APTO project under contract number 2403999. 
The author thanks F. Verastegui-Grunewald, J. Doyne Farmer and F. Lafond for their helpful comments.

\appendix
\section*{Appendix} \label{sec:Appendix}
\section{Materials and methods} \label{apx:materials}

\begin{table}[ht]
\centering
\begin{tabular}{|l|r|l|}
  \hline
& Nr. unique & Comments \\ 
& observations &  \\ 
  \hline
Years &  23 & 2001-2023 \\ 
  Assets & 146,720 & i.e., plant units \\ 
  Power plants & 72,739 &  \\ 
  Direct owners & 30,761 & Appendix~\ref{apx:owner_map} \\ 
  Firms (ultimate owners) & 23,196 & Appendix~\ref{apx:owner_map} \\ 
  Tech. categories &  12 & Appendix~\ref{apx:tech_map} \\ 
  Firm-asset pairs & 217,195 &  \\ 
  Firm-tech. pairs & 31,385 &  \\ 
  Firm-tech.-year triples & 490,941 &  \\ 
  \# Investments & 147,477 & $\Delta \text{MW} >0^*$\\ 
  \# Retirements & 50,754 & $\Delta \text{MW} <0^*$\\ 
   \hline
\end{tabular}
\caption{ {\bf Data overview.}
$^*$ computed at the asset-firm-year level.}
\label{tab:dataview}
\end{table}

\subsection{Data processing} \label{apx:data}
Our main data source is S\&P Capital IQ Pro -- Energy (CIQ).
This database contains extensive information on power plants and individual power plant units (e.g., individual turbines within a power plant), including nameplate capacities in MW, detailed geographical coordinates and the underlying technology.
Moreover, we know in which year (and typically in which month) a plant went online for the first time and when it was retired (if it was retired). 
Due to data quality reasons, we restrict our analysis to the time horizon between 2001 and 2023.

Importantly, we also have access to the detailed ownership structure of assets, i.e., relative stakes in an asset owned by legal entities and how they change over time. For simplicity, we call these legal entities firms, although we acknowledge that a minority of them are non-firm entities. 
CIQ provides extensive firm information, allowing us to analyze owners with respect to their industry classification, headquarter locations or establishment dates. Moreover, we know the publicly owned shares of a firm and have detailed information on a firm's parents and ultimate parents (see Appendix~\ref{apx:owner_map}).

As shown in Table~\ref{tab:dataview}, this database contains almost 150k unique power plant units (which we call assets) that are part of more than 70k power plants and owned by about 30k legal entities.
In many cases, power plant units are owned by multiple owners. In such cases, our database provides detailed information on the ownership structure. For example, if multiple owners hold an asset, we can infer the name and unique identifiers of the owners as well as their percentage stake in the asset. In 2023, we observe that about 15\% of assets in our sample have multiple owners.
We aggregate asset-level capacities to owner portfolios by taking relative ownership shares into account. For example, if a 10 MW hydro plant is 60\%-owned by Owner 1 and 40\%-owned by Owner 2, we record that Owner 1 holds 6 MW and Owner 2 holds 4 MW of hydropower in their portfolios.

\subsection{Ownership mapping} \label{apx:owner_map}

Power assets are frequently owned by firms only indirectly via subsidiaries. To develop plants of a technology that was not previously held, a company may launch a new specialized subsidiary instead of integrating these assets into its existing portfolio. 
For example, the state-owned Norwegian oil and gas company Equinor ASA has substantially invested in offshore wind through its UK-based subsidiary Equinor New Energy Ltd.
Focusing only on direct asset ownership would miss such cases. 
Thus, we conduct our analysis at the level of ultimate parents, representing the highest-level entity within a corporate structure that exercises control over the subsidiaries. (We find that about 44\% of capacity is owned through operating subsidiaries.)

While CIQ provides data on ultimate parents of direct owners, we are not using this information directly because ultimate parents frequently refer to majority-holding countries or states. To give a concrete example, the largest Austrian electric utility, Verbund AG, is majority-owned by the Republic of Austria. Consequently, the Republic of Austria is the ultimate parent of Verbund AG and all its subsidiaries. Since the relevant strategic business decisions arguably take place at the level of Verbund AG, we aim to map all subsidiaries' assets to Verbund AG rather than to the Republic of Austria.

Because our data further provides information on the direct parents of firms, a simple alternative approach would be to analyze the power asset portfolios at the level of the direct owners' parents. 
However, this approach neglects the possibility of more complex ownership relationships where the parent of a subsidiary could be again a subsidiary of another company. 
Consequently, we adopt a custom strategy to map direct asset owners to their ultimate company parents while avoiding mapping assets to countries or states.

More specifically, we adopt a ``snowballing'' algorithm that starts at the level of direct asset owners and percolates through the branches of the ownership tree until it arrives at a leave representing an independently operating company. We do this by taking advantage of a variable in our dataset that classifies firms into \emph{operating subsidiaries} vs. \emph{operating} firms, where the latter indicates that the firm is not a subsidiary of another company. For example, Equinor ASA and Verbund AG are classified as \emph{operating}, while Equinor New Energy Ltd. and Verbund Hydro Power GmbH are classified as \emph{operating subsidiary}.
In the first step of the algorithm, we test if the direct owner is associated with the status \emph{operating} (as opposed to \emph{operating subsidiary}). If true, the direct owner is considered as the ultimate parent company, and the algorithm stops. However, if the direct owner is associated with the status \emph{operating subsidiary}, we retrieve the parent firm of the direct owner and check the status of the parent firm. If the parent firm has the status \emph{operating}, we have found the ultimate parent of the direct owner, and the algorithm stops. However, if the parent company is again an \emph{operating subsidiary}, we go one level up in the ownership tree and consider the status of the parent of the parent. This procedure is repeated until we arrive at a leaf in the ownership tree with the status \emph{operating}. 
For some firms, multiple parents are listed. In this case, we assign the subsidiary in equal parts to its parents. For example, if three parents of a direct owner holding 9 MW of power capacity are listed, we would assign 3 MW to each of the parents.

In cases where the snowballing procedure did not lead to the identification of parents labeled as \emph{operating} (applying to owners representing about 15\% of total capacity), we used the ultimate parent label provided by CIQ. Note that we aggregated the ownership only based on parents-operating subsidiary relationships but did not aggregate the assets of portfolio companies to their investors (see SI~Section~S2). 

This approach maps about 30.7k direct owners into 23.2k ultimate parents, which we refer to as firms or companies in the main text. Overall, this procedure results in more than 200k unique asset-firm ownership observations (Table~\ref{tab:dataview}).

\subsection{Technology aggregation} \label{apx:tech_map}

Our plant units are associated with detailed technological categories and fuel groups, which we aggregate into twelve electricity-generating technology categories: 
solar PV, solar thermal, wind offshore, wind onshore, biomass, hydro (including pumped-hydro storage), geothermal, waste, nuclear, oil, gas and coal.
Doing this results in 31.4k unique observations of positive firm-technology capacity holdings, and almost 500k observations of firm-technology-year triples (Table\ref{tab:dataview}).

In the main text, fossil-based plants refer to plants that are based on oil, gas or coal fuels. Plants based on solar radiation, wind, biofuels, geothermal and hydropower make up the group of renewables.

\subsection{Firm transition definition} \label{apx:trans}
We consider a firm to have transitioned from fossil to renewables if the majority of its asset portfolio has shifted from fossil-fuel-based to renewable energy technologies. To test whether this condition holds for a given firm, we require that the portfolio share of renewables is >50\% in every year in the most recent 3-year period of our sample (2021--2023). 
Considering this recent multi-year period for the definition of transitioning firms avoids including firms that temporally held a renewables majority in the past but have not persistently transformed their portfolio.
Similarly, to qualify as a transitioning firm, we require that a firm held a majority in fossil assets for at least five consecutive years. This condition ensures the exclusion of ``noisy'' portfolios (e.g., portfolios switching over time between zero and small positive fossil capacities).

We acknowledge that other definitions could be considered but emphasize that any definition must make specific subjective choices regarding portfolio share thresholds and time periods. Our transition definition is not particularly strict. In an extreme case, a transitioned portfolio could still consist of 49.9\% of highly carbon-intensive assets. However, even with this definition, only a small group of firms have transitioned. To understand the robustness of our approach, we have explicitly explored the impact of choosing alternative thresholds and extensively discussed results when considering a 75\% renewable portfolio share threshold (Sections~\ref{sec:firm_trans}-\ref{sec:tech_trans}).

\subsection{Data coverage} \label{apx:coverage}

According to CIQ, our sample covers essentially all plants
that file data with the U.S. Energy Information Administration or are larger than 1 MW in North America and 5 MW outside North America.
The data also comprises a number of plants that fall below these thresholds. For example, about 1/3 of all plants outside of North America in our sample are smaller than 5 MW. Overall, we observe plants in 220 countries (this number is higher than the actual list of sovereign states since specific non-independent territories are considered as states as well; e.g., Falkland Islands, Faroe Islands, Cayman Islands, Montserrat). One notable exception is Russia, for which no power plant data is included (Russian firms holding power assets abroad are included, though).
In SI~Section~S1, we provide further detail on sample size and compare the coverage of our data to alternative estimates of global installed power capacities.

\subsection{Data availability}
The data on power-generation assets and owners have been obtained under an S\&P Global Market Intelligence license and, therefore, cannot be made public.

\subsection{Code availability}
The analysis was conducted in R. Since the underlying data cannot be made public, the code was not published but is available from the author upon reasonable request.

\bibliographystyle{abbrvnat}
\bibliography{references}


\end{document}


\begin{center}
\huge{\textbf{Supplementary Information}}
\end{center}
\vspace{0.05cm}

\begin{center}
\LARGE{Transition dynamics of electricity asset-owning firms}
\end{center}
\vspace{0.05cm}

\begin{center}
\large{Anton Pichler} \\[.4em]
\small Vienna University of Economics and Business\\[.1em]
\small Macrocosm Inc.\\[1em]
\large \today 
\end{center}

\tableofcontents
\newpage

\section{Data coverage} \label{si:data}

\begin{figure}[H]
    \centering
    \includegraphics[width = \textwidth]{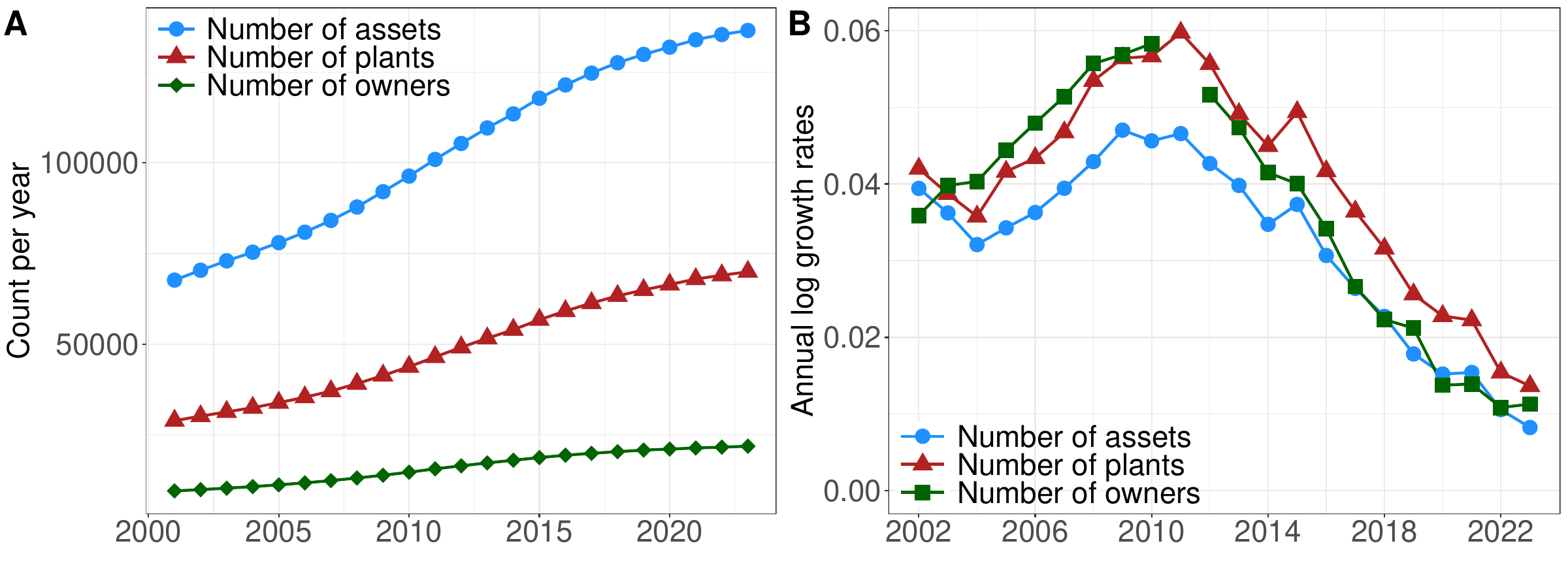}
    \caption{
    {\bf The dynamics of firm, asset, and plant numbers in our dataset.} 
    A: The total number of assets, plants and asset-owning firms. 
    B: Annual log growth rates for the number of assets, plants and asset-owning firms. 
    }
    \label{fig:apx_data_time}
\end{figure}

According to the data provider, our sample covers essentially all plants installed in North America if they are greater than 1 MW and in the rest of the world if they are greater than 5 MW. 
Fig.~\ref{fig:apx_data_time}A shows the number of owners (i.e., at the custom ultimate parent level as described in the main text's appendix), the number of plants, and the number of assets over time. We show the corresponding annual log growth rates in Fig.~\ref{fig:apx_data_time}B. In our database, the unique numbers of assets, plants, and owners have increased steadily over time, although at a lower rate in recent years.

\begin{figure}[H]
    \centering
    \includegraphics[width = \textwidth]{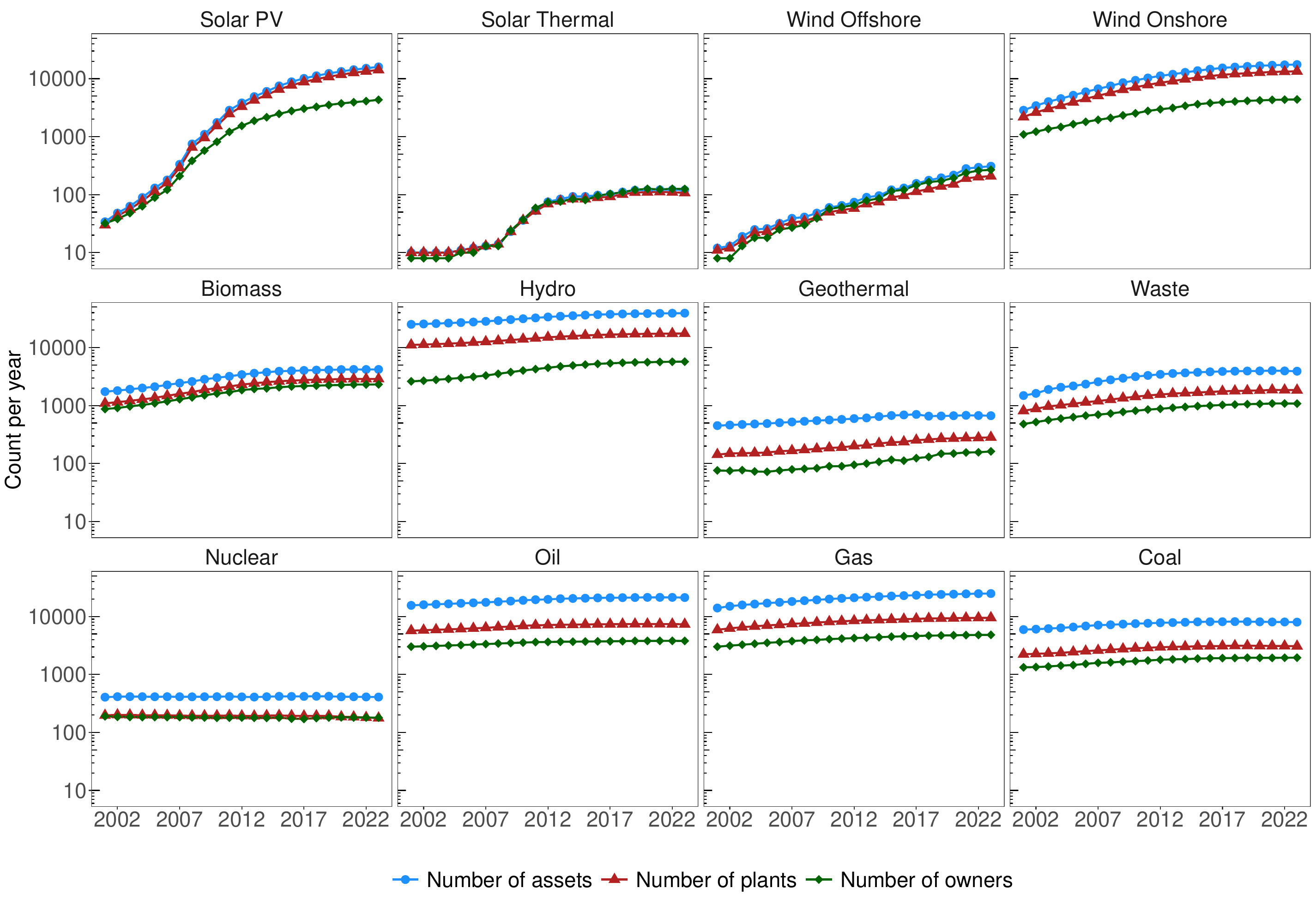}
    \caption{{\bf Technology-specific descriptive statistics of the asset-level dataset.} The figure shows the number of assets (power plant units), number of power plants and number of owners per technology per year. The y-axis is on log-scale.
    }
    \label{fig:apx_nr_ass_year}
    \centering
    \includegraphics[width = \textwidth]{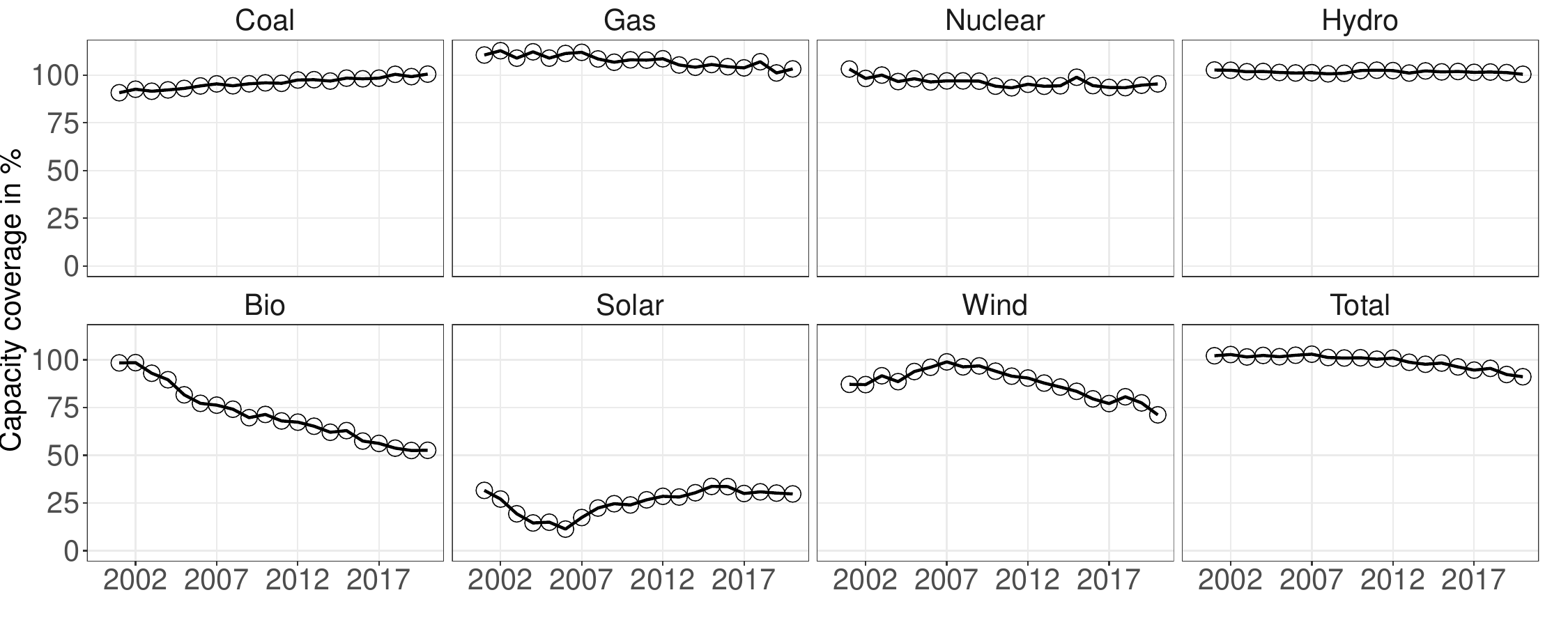}
    \caption{{\bf Data coverage per aggregate technology over time.} Capacity coverage per year of our asset-level dataset when compared to aggregate statistics as reported by Ember.}
    \label{fig:apx_data_cov}
\end{figure}

In Fig.~\ref{fig:apx_nr_ass_year}, we show the annual numbers of assets, plants and owners broken down for each individual technology. The figure shows that the numbers of assets and firms are highly unevenly distributed over technologies. For example, we observe about 10k unique hydro, gas, and oil plants, while the numbers of solar thermal and offshore wind plants lie in the hundreds. For some technologies, such as solar PV, offshore wind, and onshore wind, we find a substantial increase in assets and firms owning them. 
Note that we also observe many nuclear plants and their owners in our dataset. The World Nuclear Association reports 436 nuclear reactors in 2023\footnote{
\url{https://www.world-nuclear.org/information-library/facts-and-figures/reactor-database.aspx} (accessed: 2023/11/15)
}, compared to 408 nuclear reactors present in our sample in 2023.

To check the coverage of our asset-level dataset, we compare total technology-specific capacities in our sample with the global capacity statistics reported by Ember \citep{ember2022}. To do this, we aggregate our asset data into technology categories used by Ember and divide the total technology-specific capacity in our data by the total global capacity reported by Ember. Fig.~\ref{fig:apx_data_cov} shows that our data has high total coverage, although we observe some differences across technologies. 
For some years and technologies (e.g., hydro and gas), the asset-level capacities reported in our data sum up to higher numbers than those reported by Ember. This is likely due to uncertainties in reported power capacities, whether a plant was online or offline in a given year, and differences in how power plants can be categorized into technological groups. 
While we observe excellent coverage for most technologies, there are a few exceptions. For example, we have essentially full coverage for biopower in the early years of our sample, but it decreases substantially over time. Similarly, we have relatively good coverage for wind power (on- and offshore) but also observe a decline in coverage over time. The coverage is lower for solar (PV and thermal), which lies around 25\% in recent years. The lower coverage for solar is expected, given that a significant share of solar PV plants is small-scale.
For example, the International Energy Agency reports that only about 2/3 of total solar PV capacities are ``utility-scale''\footnote{
\url{https://www.iea.org/reports/renewables-2020/solar-pv} (accessed: 2023/12/21).
}.
Moreover, even if utility-scale, we expect many solar plants to fall below our data's capacity thresholds (1 MW in North America and 5 MW else).

\newpage
\section{Operating subsidiaries and portfolio companies}
\label{sec:firm_map}
CIQ distinguishes between different types of firm ownership. In particular, it distinguishes between portfolio companies and subsidiary companies. 
A subsidiary company is controlled by another company, known as the parent company, through majority ownership of its shares. Subsidiaries can operate independently but are still part of the larger corporate structure of the parent.
In contrast, a portfolio company typically refers to a company that is part of a private equity or venture capital firm's investment portfolio. These companies are often not wholly owned but are significant investments for the firm. They operate independently and are regarded as independent of the larger corporate structure of the parent. Consequently, they are typically classified as \emph{operating} as opposed to \emph{operating subsidiary}.

The algorithm described in the appendix of the main text is built on the classification of \emph{operating} vs. \emph{operating subsidiaries}. As a result, our aggregation does not take asset-ownership through portfolio companies into account but only if operating subsidiaries hold them. 

Let us illustrate this with a concrete example that becomes apparent from Table~\ref{tab:top25}. Huadian Power International Corporation Limited (rank 18 in Table~\ref{tab:top25}) is 44\% owned by China Huadian Corporation Ltd (rank 8). Huadian Power International is classified as \emph{operating}, and no parent with majority ownership exists. Consequently, we treat the firm as an independent entity and do not further aggregate its assets to its owners. Clearly, alternative choices could be made, given that Huadian Power International is generally considered to be under the effective control of China Huadian Corporation despite owning less than 50\% of its shares. Similar reasoning applies to Datang International Power (rank 14) and China Datan Corporation (rank 9).

\newpage
\section{Further results}

\subsection{Ownership based on industry types} \label{sec:industry_own}

We use firm industry codes provided by CIQ to analyze industry-specific technology ownership. Of the more than 23k firms in our sample, 18k are associated with a primary industry category. In total, we observe 162 unique S\&P custom industry categories. We aggregate these categories in broader industry classes. For example, we observe several distinct utility classes (the vast majority of them are electric utilities, but there are also others, such as water or gas utilities). We group them all into a single utility class. CIQ also has separate industry codes for independent power producers (IPP) and renewable energy providers. We group both categories into an IPP class. We further group all (non-utility) firms in the gas, oil and coal sectors into a single industry class. 

\begin{figure}[ht]
    \centering
    \includegraphics[width=\linewidth]{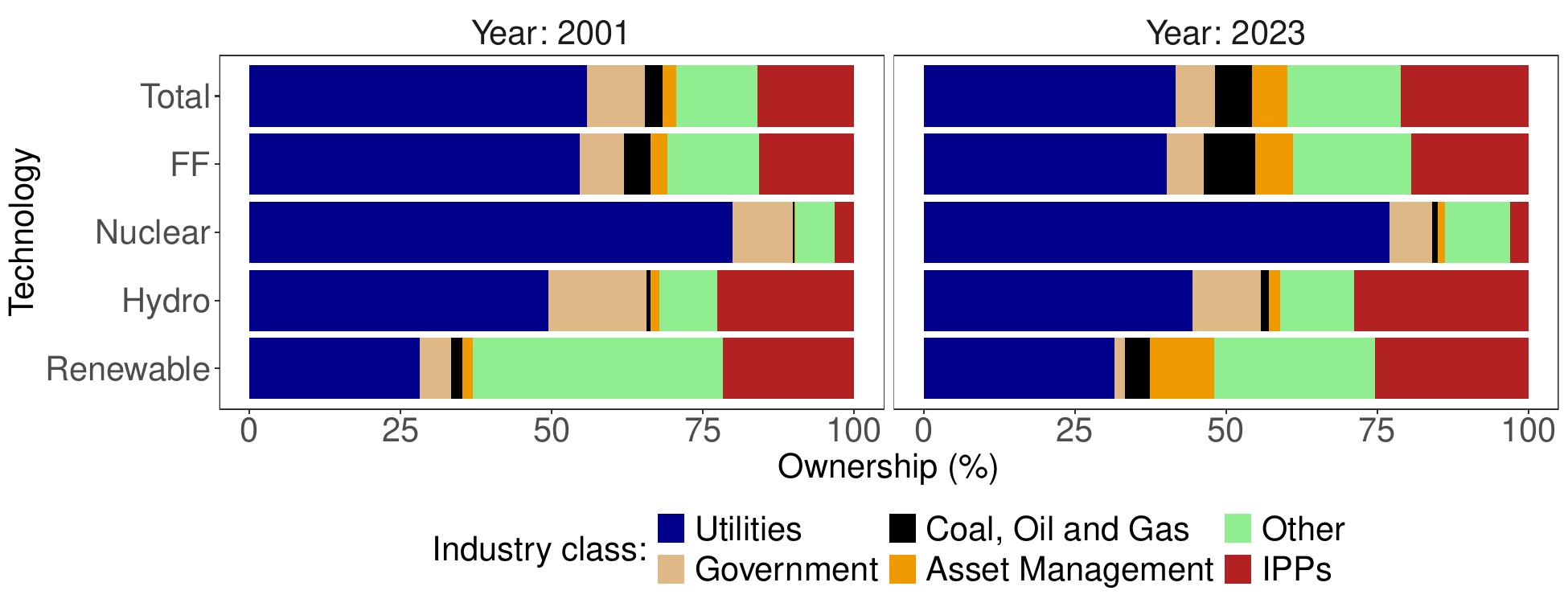}
    \caption{
    {\bf Technology ownership based on firm's industry classes.}
    Renewable refers to non-hydro renewable. The class \emph{Other} refers to firms that are not classified at all or classified with an industry category not shown here. 
    }
    \label{fig:industry_own}
\end{figure}

We visualize the ownership split across the top-owning industry classes for 2001 and 2023 in Fig.~\ref{fig:industry_own}. 
In our global sample, firms classified as utilities own about 42\% of total installed capacity in 2023 compared to 56\% in 2001. The second most important group of owners are IPPs, which own about 21\% in 2023 compared to 16\% in 2021. 
Government institutions, asset management firms and coal, oil and gas firms each own about 6\% of assets in 2023. The remaining ownership is distributed across firms with varying industry associations. 

Utilities dominate nuclear power ownership and are also the largest owners of fossil and hydropower, while they hold a comparatively small share of non-hydro renewables. IPPs and firms grouped into diverse industry categories own the majority of non-hydro-renewable assets. Asset management firms play an increasingly important role, owning about 10\% of total non-hydro-renewable capacities in 2023.

\newpage

\subsection{Ownership distribution} \label{sec:size}

Electricity generation capacities are unevenly distributed across firms. While the top 25 firms represent less than 0.1\% of all firms in our sample, they own almost 2,000 GW of power capacity, representing about a quarter of the total global installed capacity (Table~\ref{tab:top25}).\footnote{
Due to the imperfect coverage of (small) plants in our sample, the share of global capacity of the top 25 firms is likely somewhat below this number.
}. 
Many of the largest firms are located in Asia, with ten of the largest power firms headquartered in China alone. The top 25 companies own a disproportionate share of coal, accounting for 40\% of global installed coal capacity. Conversely, gas power is underrepresented in their portfolios when compared to the global average.
Considering that reducing coal usage is a key pillar in any net zero pathway (e.g., \cite{IEA2021netzero}), large firms will likely be disproportionately affected by the decarbonization of the power sector.

Thus, power capacities are unequally distributed over firms. In Table~\ref{tab:sizetab} we show that 80\% of total global power capacities are owned by 3\% of the firms in our sample. The unequal distribution of power capacities over firms is also confirmed through a large Gini index 0.94. In line with these observations, we find that the average-sized firm ($\approx$ 330 MW) is substantially larger than the median-sized firm ($\approx$ 10 MW).

\begin{figure}[H]
    \centering
    \includegraphics[width = \textwidth]{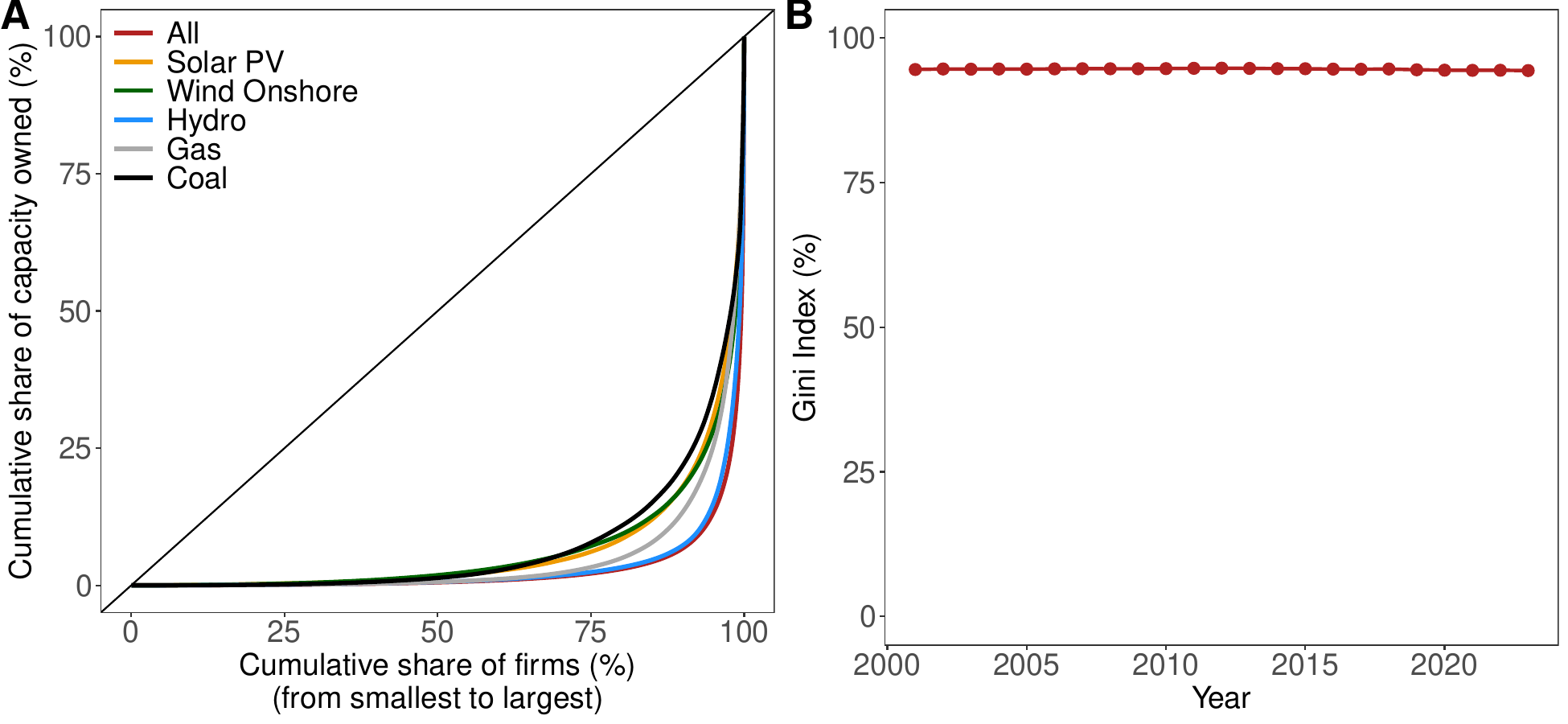}
    \caption{
    {\bf A. Lorenz curve of capacity ownership.} 
   Data is based on the year 2023.
   {\bf B. Gini index on total capacity ownership shares computed for every year.} The index varies slightly from year to year (between 0.944 and 0.948) which is hardly visible if considering the whole theoretical range of the index.
   }
    \label{fig:ecdf}
\end{figure}

The table further shows that we observe similar uneven ownership distribution in essentially every technology class, although the Gini indices for individual technologies tend to be somewhat lower. Fig.~\ref{fig:ecdf}A further emphasizes the uneven distribution for specific key technologies by visualizing the Lorenz curve of technology-specific capacity ownership. The empirical cumulative distribution of ownership shares strongly deviates from a theoretical uniform ownership distribution for all technologies.

Fig.~\ref{fig:ecdf}B shows the Gini index computed for every single year based on the total capacity ownership of firms. As the figure shows, the Gini index has been virtually constant over the past two decades.

\begin{table}[H]
\centering
\resizebox{\textwidth}{!}{%
\begin{tabular}{|rllrrrrrrrrr|}
	\hline
	& Owner & Country & Capacity & Share & Coal & Gas & Oil & Nuclear & Hydro & Solar & Wind \\ 
	& (ultimate parent)  &  & (GW) & (\%) & (\%) & (\%) & (\%) & (\%) & (\%) & (\%) & (\%) \\ \hline
	1 & CHN Energy Investment Corp & CN & 204.8 & 2.9 & 84.9 & 2.3 & 0.4 & 0.1 & 6.8 & 0.7 & 4.7 \\ 
	2 & Electricit\'e de France & FR & 125.2 & 1.8 & 4.2 & 10.2 & 3.5 & 50.7 & 18.6 & 2.9 & 9.8 \\ 
	3 & Huaneng Power Intl & CN & 104.0 & 1.5 & 82.5 & 10.3 & 0.0 & 0.4 & 1.3 & 1.4 & 4.0 \\ 
	4 & China Three Gorges Corp & CN & 99.3 & 1.4 & 0.0 & 0.0 & 0.0 & 0.0 & 97.5 & 1.0 & 1.5 \\ 
	5 & Korea Electric Power Corp & KR & 98.0 & 1.4 & 38.7 & 22.8 & 6.0 & 25.1 & 5.7 & 0.7 & 0.8 \\ 
	6 & State Power Investment Corp & CN & 93.4 & 1.3 & 55.6 & 5.1 & 0.0 & 6.9 & 16.6 & 6.5 & 9.0 \\ 
	7 & Enel SpA & IT & 88.3 & 1.3 & 9.4 & 20.9 & 5.1 & 4.7 & 30.3 & 10.4 & 17.9 \\ 
	8 & China Huadian Corp & CN & 81.2 & 1.2 & 53.8 & 9.5 & 0.7 & 0.3 & 27.2 & 3.1 & 5.2 \\ 
	9 & China Datang Corp & CN & 74.1 & 1.1 & 56.3 & 3.7 & 0.0 & 0.0 & 27.9 & 1.0 & 11.1 \\ 
	10 & NextEra Energy Inc & US & 73.6 & 1.0 & 1.0 & 39.1 & 1.4 & 8.0 & 0.0 & 16.8 & 33.8 \\ 
	11 & NTPC Ltd & IN & 72.6 & 1.0 & 81.2 & 8.8 & 0.1 & 0.0 & 5.2 & 4.4 & 0.3 \\ 
	12 & China Huaneng Grp & CN & 70.2 & 1.0 & 50.7 & 1.6 & 0.2 & 0.6 & 35.6 & 0.9 & 10.2 \\ 
	13 & Engie SA & FR & 68.8 & 1.0 & 3.3 & 43.9 & 3.3 & 7.1 & 21.9 & 5.9 & 13.7 \\ 
	14 & Datang Intl Power & CN & 66.7 & 0.9 & 77.2 & 8.3 & 0.0 & 2.8 & 9.6 & 0.4 & 1.6 \\ 
	15 & Iberdrola SA & ES & 60.3 & 0.9 & 0.0 & 27.6 & 0.5 & 5.0 & 23.9 & 6.0 & 36.9 \\ 
	16 & Duke Energy Corp & US & 59.8 & 0.8 & 19.0 & 46.1 & 2.9 & 15.4 & 6.3 & 6.2 & 4.2 \\ 
	17 & Egyptian Ele. Holding Comp & EG & 57.9 & 0.8 & 0.0 & 91.6 & 3.0 & 0.0 & 5.4 & 0.0 & 0.0 \\ 
	18 & Huadian Power Intl & CN & 57.5 & 0.8 & 78.2 & 11.1 & 0.0 & 0.0 & 3.3 & 0.2 & 7.3 \\ 
	19 & China Resources Comp & CN & 56.4 & 0.8 & 89.8 & 1.0 & 0.3 & 0.0 & 0.7 & 0.1 & 8.0 \\ 
	20 & TEPCO & JP & 53.8 & 0.8 & 12.6 & 44.3 & 5.6 & 16.4 & 18.5 & 0.3 & 2.3 \\ 
	21 & RWE AG & DE & 52.6 & 0.7 & 22.8 & 30.6 & 0.7 & 2.8 & 6.6 & 15.0 & 21.1 \\ 
	22 & Eskom Holdings & ZA & 52.1 & 0.7 & 85.1 & 0.0 & 4.6 & 3.6 & 6.5 & 0.0 & 0.2 \\ 
	23 & Electrobas & BR & 51.1 & 0.7 & 0.7 & 2.4 & 3.1 & 1.3 & 91.0 & 0.0 & 1.4 \\ 
	24 & Public Investment Fund (PIF) & SA & 49.6 & 0.7 & 0.0 & 39.6 & 59.9 & 0.0 & 0.0 & 0.5 & 0.0 \\ 
	25 & Comisi\'on Federal de Ele. (CFE) & MX & 49.5 & 0.7 & 7.9 & 34.5 & 24.6 & 3.3 & 27.3 & 0.1 & 0.4 \\ \hline
	& Total (Top 25) & ~ & 1921 & 27.2 & 40.2 & 17.6 & 3.8 & 7.2 & 19.6 & 3.3 & 8.1 \\ 
	& Total (All) & ~ & 7055 & 76.1 & 30.1 & 26.9 & 5.4 & 5.4 & 18.2 & 4.5 & 8.3 \\ \hline
\end{tabular}
}
\caption{{\bf Top 25 electricity-asset owning companies in 2023.}  \emph{Capacity} shows total power capacities owned in GW and \emph{Share} presents the firms' share of global power capacity. Columns referring to technological categories indicate the share of that technology in a given portfolio. \emph{Solar} considers both solar PV and solar thermal and \emph{Wind} considers both on- and offshore wind power. 
Note that capacities shown here can differ from numbers shown in other sources since assets can be aggregated differently through a company's corporate structure (see Section~\ref{sec:firm_map}). 
}
\label{tab:top25}
\centering
\vspace{.4cm}
  \resizebox{\textwidth}{!}{%
\begin{tabular}{|l|rrrrrrrrrrrrr|}
  \hline
 & All & Solar & Solar & Off- & On- & Bio & Hydro & Geo & Waste & Nuclear & Oil & Gas & Coal \\ 
  &  &  PV & thermal  & shore & shore &  &  &  &  &  &  &  &  \\ 
  \hline
Capacity owned by top 0.1\% & 25.3 & 12.4 & 12.4 & 6.8 & 15.3 & 9.4 & 22.2 & 8.4 & 8.5 & 16.6 & 19.9 & 10.0 & 14.9 \\ 
Capacity owned by top 1\% & 63.0 & 38.2 & 22.8 & 22.1 & 44.9 & 30.8 & 55.7 & 20.6 & 23.8 & 29.0 & 46.4 & 39.4 & 40.1 \\ 
Capacity owned by top 5\% & 86.7 & 69.3 & 44.0 & 49.0 & 71.7 & 56.1 & 85.0 & 47.9 & 48.5 & 51.5 & 74.2 & 74.5 & 65.1 \\ 
80\% of capacity owned by & 2.9 & 8.9 & 27.6 & 20.5 & 8.6 & 18.6 & 3.7 & 17.5 & 22.4 & 20.2 & 7.6 & 6.8 & 11.2 \\ 
Gini index & 94.4 & 87.4 & 70.7 & 77.2 & 87.4 & 78.1 & 93.6 & 78.9 & 74.1 & 77.7 & 88.7 & 90.5 & 86.0 \\ 
  Observations & 21625 & 4246 & 123 & 264 & 4336 & 2284 & 5673 & 160 & 1072 & 178 & 3771 & 4765 & 1937 \\ 
   \hline
\end{tabular} }
\caption{ {\bf Firm size distribution summary.}
All values are given in \%, except for the last row, which shows the number of observations for each column, i.e., counts of firms (owners). 
Data is based on the year 2023.
}
\label{tab:sizetab}
\end{table}

\newpage
\subsection{Size, firm focus, and number of technologies owned}

\begin{figure}[H]
    \centering
\includegraphics[width = \textwidth]{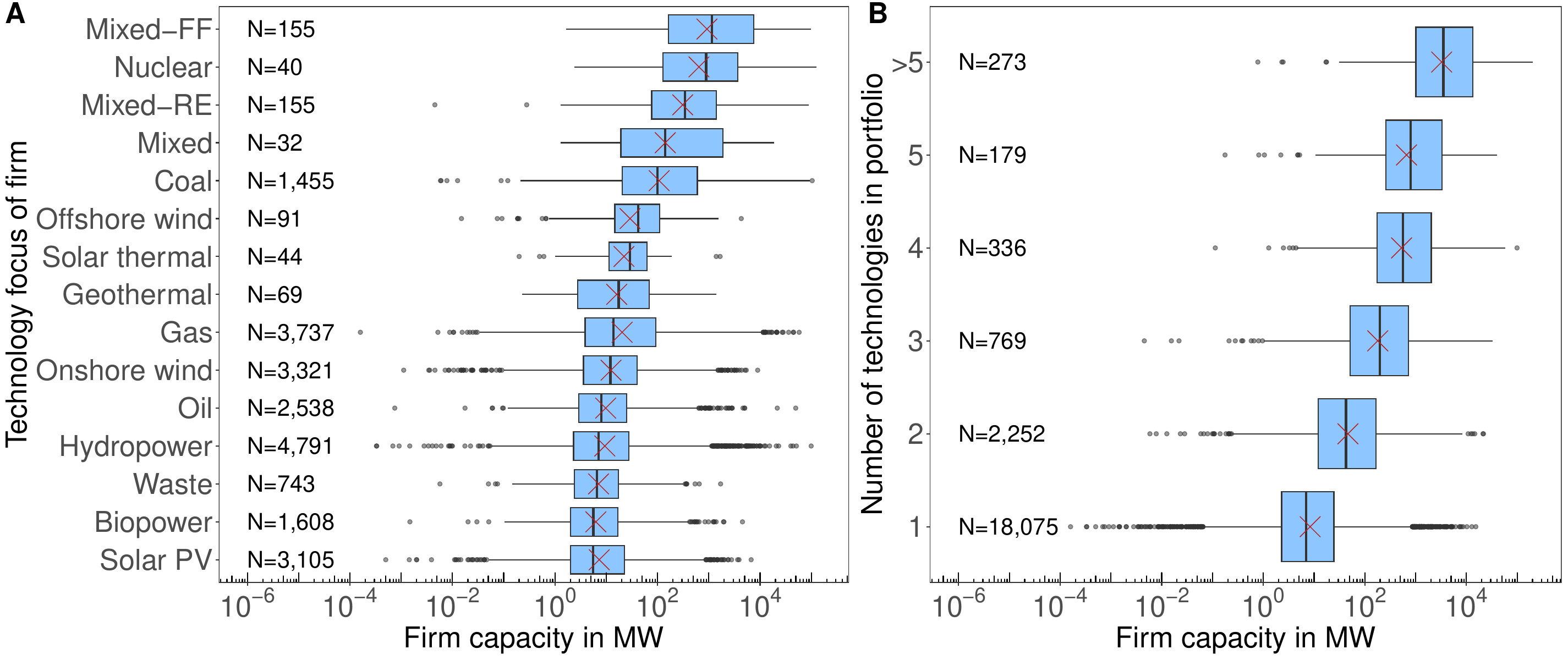}
    \caption{
    {\bf A. Firm total capacity by technology focus.}
The technology focus of a firm is defined as the technology that accounts for more than 50\% of the firm’s capacity. If no single technology exceeds 50\%, but the combined capacity of all fossil technologies makes up more than 50\%, the firm is classified as \emph{mixed-FF}. Similarly, \emph{mixed-RE} firms are those where renewable technologies (including waste but excluding nuclear) together account for more than 50\% but no single technology represents a majority. Firms that do not meet either of these criteria are classified as \emph{mixed}.
   {\bf B. Firm total capacity by number of technologies owned.}
In both panels, red crosses indicate averages and $N$ represents the number of firms per group. Both panels are based on 2023 data. Note that the x-axes are on log scale.
}
    \label{fig:focsize}
\end{figure}

In Fig.~\ref{fig:focsize} we show capacity distributions for firms belonging to different groups of technological specialization (Panel A) and the number of technologies held (Panel B). Firms focusing on solar PV have the smallest median capacity value. Conversely, mixed-FF firms -- those with a diversified fossil portfolio where no single fossil technology dominates but together constitute a majority -- have the largest median capacity. Note that the difference between the two groups is enormous. While the median solar PV firm holds 5.5 MW total capacity, the median mixed-FF firm capacity is 1,150 MW, corresponding to a size difference by a factor of 200.
The spread of the size distribution can be substantial for some technology focus groups.  
For example, the solar-dominated Indian company Adani Green Energy holds almost 7,000 MW of power capacity.
We can find examples of small and extremely large firms across most technology focus groups.

Fig.~\ref{fig:focsize}B shows that multiple-technology firms tend to be large but represent a minority. Only about $1\%$ of firms in our sample own more than five technologies. Nevertheless, they represent approximately 50\% of global capacities in 2023 (this share has increased substantially over time). For comparison, while more than 80\% of firms are single technology firms, they only account for about 14\% of global power capacity. 
We observe that the average single technology firm is by a factor of 230 smaller than the average firm holding more than five technologies.

While a positive correlation between the number of technologies held in a portfolio and total firm capacity may be expected, it is noteworthy that the size distribution of firms owning only one technology is spread out widely, i.e., we observe very small \emph{and} very large firms that hold only a single technology. Overall, we find 166 single technology firms larger than 1 GW.
Conversely, we also find relatively small firms that own less than 10 MW but hold several technologies in their portfolio. Thus, technological diversification strongly correlates with firm size but there are remarkable exceptions, as demonstrated by the existence of large but highly focused and small but diversified energy asset-owning firms.

\newpage
\subsection{Technology concentration}

To quantify concentration in technology portfolios, we focus on the share of the largest technology in the portfolio. We have considered using alternative ``standard'' concentration measures, such as the Herfindahl index. However, due to the small sample sizes involved (there are at most twelve technologies in a portfolio), the Herfindahl index or the normalized Herfindahl index becomes hard to interpret, in particular, if we compare values for portfolios with different numbers of technologies.

\begin{figure}[H]
    \centering
\includegraphics[width = .95\textwidth]{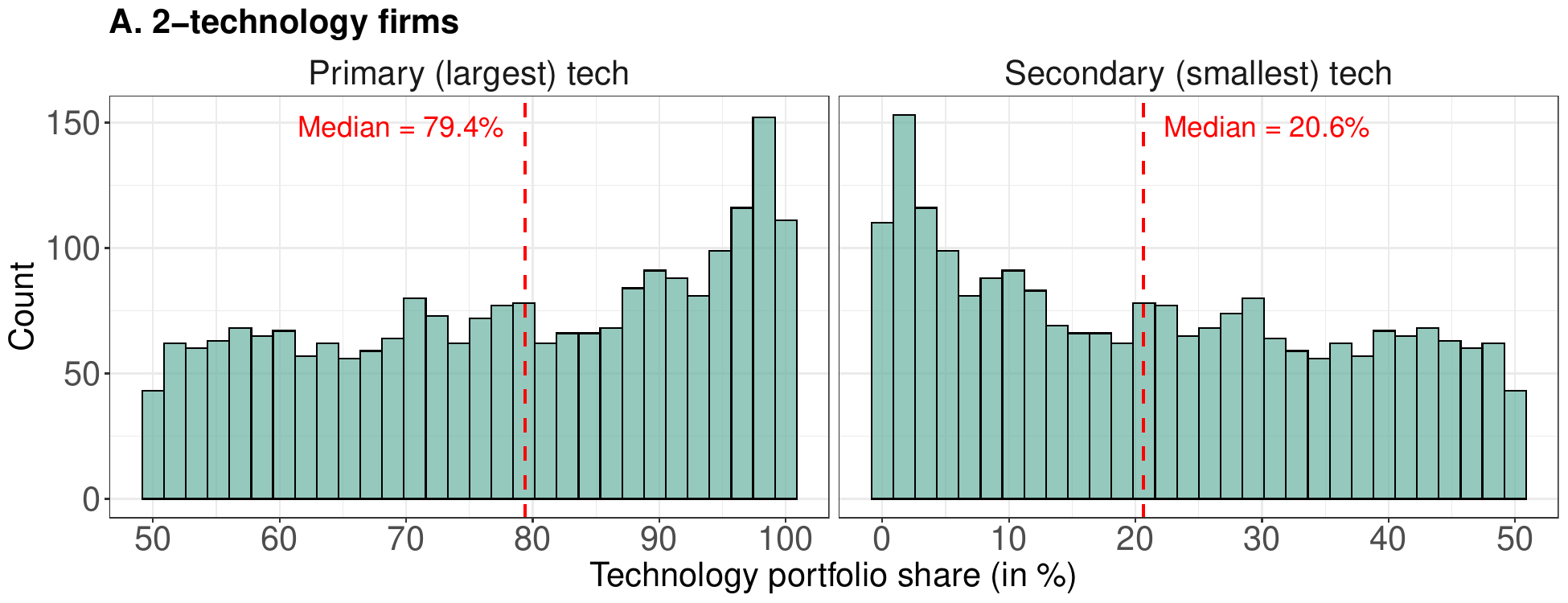}
\includegraphics[width = .95\textwidth]{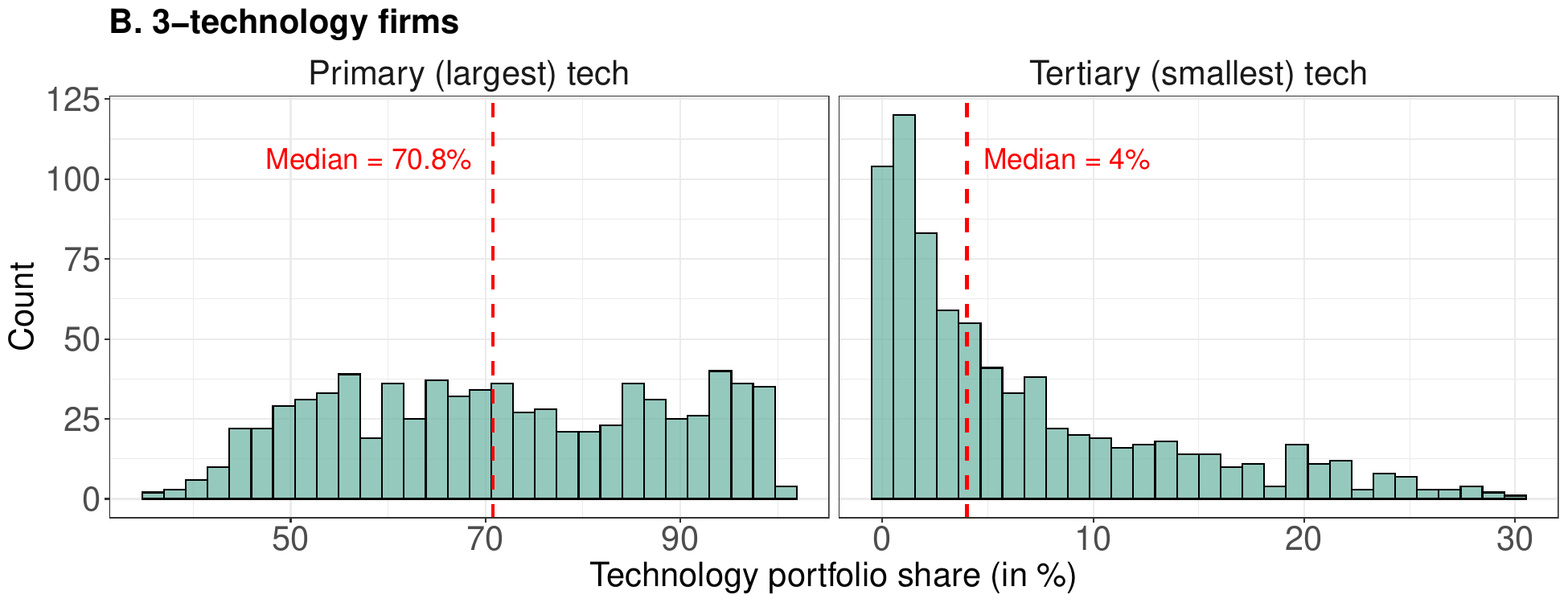}
\includegraphics[width = .95\textwidth]{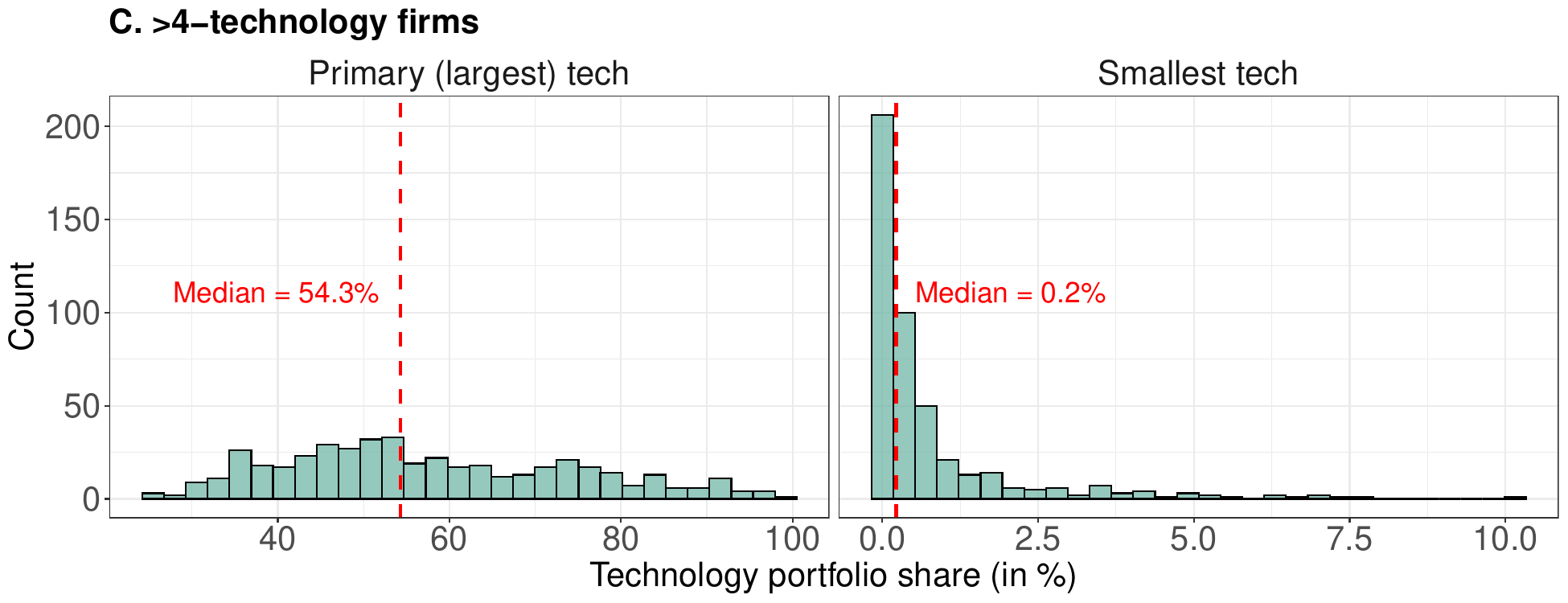}
    \caption{
    {\bf Distribution of portfolio shares of the largest and smallest technology class for multiple-technology firms.}
All panels are based on 2023 data.
}
    \label{fig:concentr}
\end{figure}

In Fig.~\ref{fig:concentr} we show the distribution of shares of the largest and smallest technologies in the portfolios of multiple-technology firms. We see that the median portfolio of a 2-technology firm exhibits roughly an 80/20 capacity split rather than a 50/50 split. We also find high levels of technology specialization for firms owning more than two technologies. In all cases, we observe that only relatively few firms hold well-balanced technology portfolios.

\subsection{Inclination to invest}
We investigate the firm characteristics that can help explain whether a firm invests in a given year. We do this by fitting a logistic regression where the dependent variable is the binary variable indicating whether a firm invests in a specific year, and the covariates are firm-specific variables.
All covariates are lagged by one year; i.e., we model a firm's likelihood to invest in year $t+1$ based on the covariates obtained from year $t$.
In the model, we include the constant, dummies for every year (up to $t$) and the total firm log capacity. We further include four different categorical variables: the number of technologies owned (in buckets), the firm type (private, public, other), how frequently a firm has invested in the previous five years, and the firm technology focus. 

We have also experimented with controlling for firm financials (e.g., revenue, net income, return on assets, etc.). However, since we have much smaller coverage for the financial variables, we have only included ``physical'' firm characteristics. In previous specifications, we have also included another categorical variable indicating whether a firm is majority state-owned. Yet we did not find any significant contribution of this covariate and have therefore omitted it from the analysis. For a more detailed analysis of state ownership and renewable technology adoption, see \cite{BENOIT2022131796, steffen2022state}.

We fit the logistic regression model to three different time horizons (all years, the past decade, and the past five years) and report the \emph{marginal effects} (as opposed to log-odds) in Table~\ref{t:reg_invest}. 
The few firm characteristics considered here explain about one-third of the variance of whether a firm will invest in a given year.
Larger firms tend to be more likely to invest: An additional 1\% increase in capacity is associated with a 0.7\% increase in the probability of investing in a given year. We further find that publicly owned firms are more likely to invest than privately owned firms and non-firm entities.

\begin{figure}
    \centering
    \includegraphics[width=0.75\linewidth]{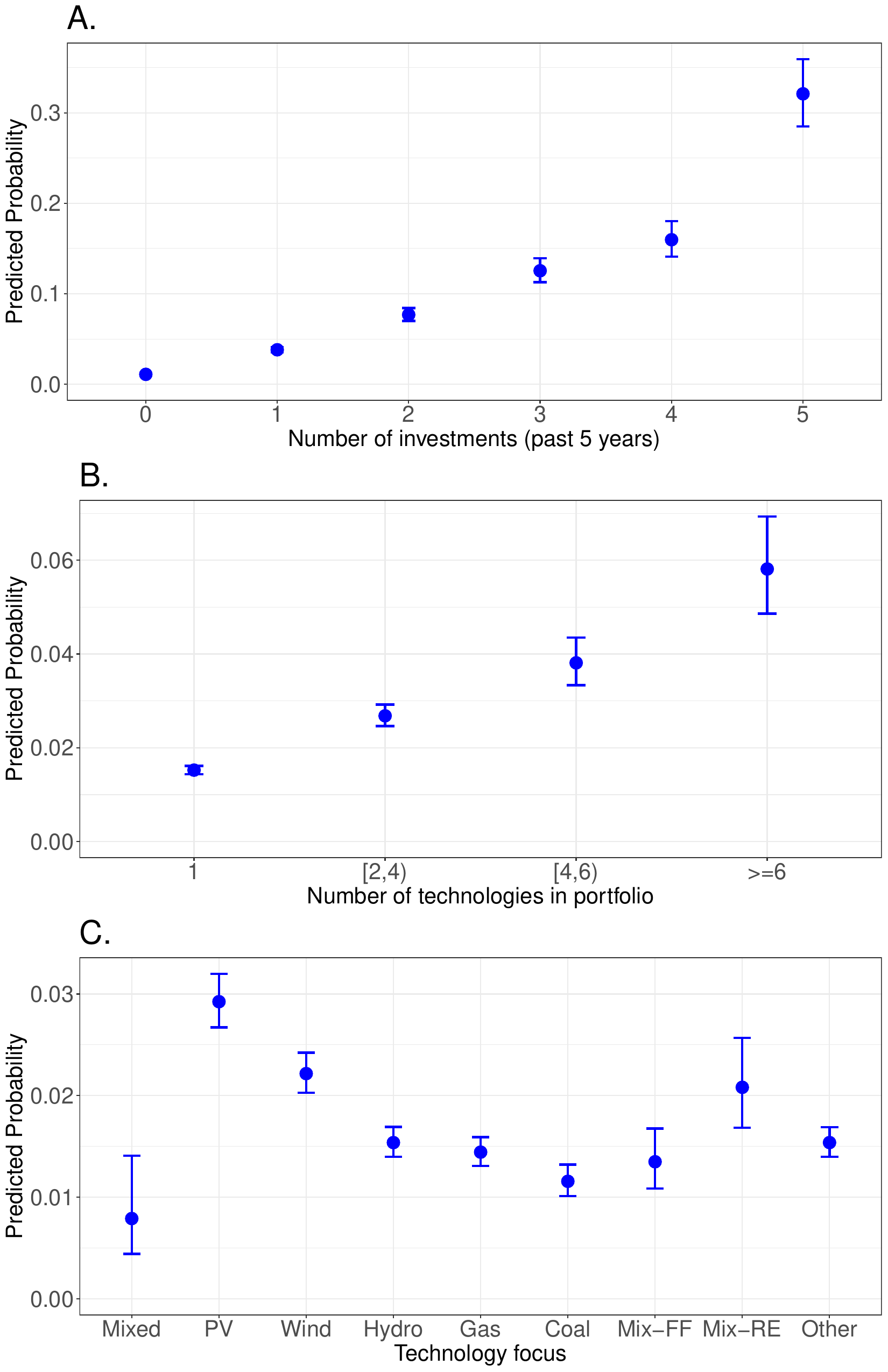}
    \caption{
    {\bf Predicted probabilities from the logistic regression model, keeping all other variables at their mean values.}
    The results presented here are based on the regression of the most recent 5-year period. Confidence bounds are computed using robust White standard errors.
    }
    \label{fig:yesno_effects}
\end{figure}

As discussed in the main text (Fig.~1), while most firms do not invest in a given year, a small group of firms invest repeatedly and consistently over time. In the logistic regression model, we find that the past investment frequency of firms is by far the most important covariate to explain its inclination to invest in subsequent years. Reformulating the model to include this variable as the sole predictor still results in a Pseudo R$^2$ of about 30\%. The probability of a firm investing in a given year increases significantly with the number of investments it has made in previous years. 
A firm that has invested only once in the previous five years is 5-6\% more likely to invest in the following year compared to a firm with no investments during that period. In contrast, a firm that has consistently invested every year is 37-50\% more likely to invest in the subsequent year than one with no investments in the past five years. 
In Fig.~\ref{fig:yesno_effects}A, we show the predicted probabilities associated with this categorical variable while controlling for all other variables.\footnote{
Note that predicted probabilities for each variable can be obtained as
$$Prob(Y=1|X) = \left(e^{\beta_0 + \sum_{i=1}^k \beta_iX_i}\right)\left(1 + e^{\beta_0 + \sum_{i=1}^k \beta_iX_i}\right)^{-1},$$ 
where $\beta_i$ represents the estimated coefficients and $X_i$ the covariates.
} The figure illustrates that the predicted probability of investing increases linearly with the number of investments up to four investments in the past five years. However, if a firm has invested in every year over the past 5-year period, its probability to invest in the following year increases disproportionately.

Another significant variable in explaining whether a firm invests in the following year is the number of technologies in a firm's portfolio. The more technologies in a portfolio, the more likely a firm is to add new power capacity. This relationship is relatively linear (Fig.~\ref{fig:yesno_effects}B).

We further find that the investment probability of a firm tends to increase if it is focused on non-hydro renewables. As in the main text, we define a firm to focus on technology $k$ if the capacity of technology $k$ accounts for more than 50\% of the firm's total capacity. If a firm holds no majority in a given technology class but a majority of all fossil capacity (i.e., when summing oil, gas and coal capacity), we define its focus as ``Mixed-FF''. We define ``Mixed-RE'' firms analogously based on renewable technologies. Consequently, ``Mixed'' firms are firms with diversified portfolios that do not hold a majority in any of these categories.
Firms focusing on solar PV tend to be more likely to invest than other firms (Table~\ref{t:reg_invest}). Also, onshore wind firms are somewhat more likely to invest in a given year, although the effect size is smaller. 
Fig.~\ref{fig:yesno_effects}C shows the predicted probabilities of investment for firms with different technology focuses. This figure makes clear that firms without a clear technology focus (mixed firms) have the lowest inclination to invest, although there is greater uncertainty around this coefficient compared to the other technology focus groups\footnote{
It is noteworthy that many mixed firms are, in fact, relatively frequent investors, with high raw probabilities of investment when not accounting for other factors. However, these firms also tend to be very large. Controlling for firm size eliminates these effects. Consequently, mixed firms' inclination to invest is comparatively low once their size is considered.
}. While all other groups exhibit a higher average inclination to invest, most of their uncertainty bounds overlap with those of mixed firms. 

\begin{table}[!htbp] \centering 
  \footnotesize
\begin{tabular}{@{\extracolsep{5pt}}lccc} 
\\[-1.8ex]\hline 
\hline \\[-1.8ex] 
 & \multicolumn{3}{c}{\textit{Dependent variable:}} \\ 
\cline{2-4} 
\\[-1.8ex] & \multicolumn{3}{c}{Firm-specific likelihood to invest in a given year} \\ 
\\[-1.8ex] 
Data sample & All years & 2013--2022 & 2018--2022\\ 
\hline \\[-1.8ex] 
\hline \\[-1.8ex] 
Constant \& year dummies   &included &included &included \\[1.2ex] 
 Log total capacity & 0.007$^{***}$ & 0.007$^{***}$ & 0.007$^{***}$ \\ 
  & (0.0003) & (0.0003) & (0.0004) \\ 
       \hdashline
  Baseline: Nr of tech owned = 1 \\[1.2ex]
  Nr of tech owned: [2,4) & 0.040$^{***}$ & 0.025$^{***}$ & 0.018$^{**}$ \\ 
  & (0.002) & (0.002) & (0.003) \\ 
  Nr of tech owned: [4,6) & 0.054$^{***}$ & 0.037$^{***}$ & 0.032$^{***}$ \\ 
  & (0.003) & (0.003) & (0.004) \\ 
  Nr of tech owned: $\ge$6 & 0.098$^{***}$ & 0.069$^{***}$ & 0.057$^{***}$ \\ 
  & (0.006) & (0.006) & (0.007) \\ 
     \hdashline
  Baseline: Private firm \\[1.2ex]
  Public firm & 0.019$^{***}$ & 0.015$^{***}$ & 0.009$^{**}$ \\ 
  & (0.001) & (0.002) & (0.002) \\ 
  Other entity & $-$0.0003 & 0.00002 & $-$0.002 \\ 
  & (0.002) & (0.002) & (0.002) \\ 
    \hdashline
  Baseline: Invested 0x in past 5 yrs\\[1.2ex]
  Invested 1x in past 5 yrs & 0.052$^{***}$ & 0.048$^{***}$ & 0.041$^{***}$ \\ 
  & (0.001) & (0.002) & (0.002) \\ 
  Invested 2x in past 5 yrs & 0.144$^{***}$ & 0.138$^{***}$ & 0.106$^{***}$ \\ 
  & (0.004) & (0.005) & (0.006) \\ 
  Invested 3x in past 5 yrs & 0.239$^{***}$ & 0.217$^{***}$ & 0.166$^{***}$ \\ 
  & (0.006) & (0.008) & (0.010) \\ 
  Invested 4x in past 5 yrs & 0.312$^{***}$ & 0.277$^{***}$ & 0.203$^{***}$ \\ 
  & (0.009) & (0.010) & (0.013) \\ 
  Invested 5x in past 5 yrs & 0.497$^{***}$ & 0.444$^{***}$ & 0.349$^{***}$ \\ 
  & (0.012) & (0.015) & (0.020) \\ 
      \hdashline
  Baseline: Focus: Mixed firm \\[1.2ex]
  Focus: Solar PV & 0.061$^{**}$ & 0.054$^{**}$ & 0.044$^{**}$ \\ 
  & (0.013) & (0.014) & (0.017) \\ 
  Focus: Wind Onshore & 0.035$^{**}$ & 0.033$^{**}$ & 0.031$^{*}$ \\ 
  & (0.010) & (0.012) & (0.015) \\ 
  Focus: Hydro & 0.018$^{*}$ & 0.017 & 0.015 \\ 
  & (0.009) & (0.010) & (0.012) \\ 
  Focus: Gas & 0.007 & 0.013 & 0.014 \\ 
  & (0.008) & (0.010) & (0.012) \\ 
  Focus: Coal & 0.005 & 0.008 & 0.007 \\ 
  & (0.008) & (0.009) & (0.011) \\ 
  Focus: Mixed-FF & 0.005 & 0.011 & 0.012 \\ 
  & (0.008) & (0.010) & (0.013) \\ 
  Focus: Mixed-RE & 0.025$^{*}$ & 0.024$^{*}$ & 0.029 \\ 
  & (0.011) & (0.012) & (0.016) \\ 
  Focus: Any other tech & 0.009 & 0.016 & 0.017 \\ 
  & (0.008) & (0.010) & (0.012) \\ 
 \hline \\[-1.8ex] 
Pseudo R$^2$ & 0.334 & 0.358 & 0.385 \\ 
Nr of investments & 22,179 & 12,457 & 5,063 \\ 
Observations & 264,638 & 177,729 & 94,270 \\ 
\hline 
\hline \\[-1.8ex] 
\textit{Note:}  & \multicolumn{3}{r}{$^{*}$p$<$0.05; $^{**}$p$<$0.01; $^{***}$p$<$1e-16} \\ 
\end{tabular} 
  \caption{
  \textbf{Marginal effects from a logistic regression of the binary dependent variable (firm investing in a given year: Yes/No) onto a set of covariates.} 
  We report average marginal effects (computed via the \emph{R} package \emph{mfx} \citep{fernihough2014marginal}) instead of the raw coefficients of the generalized linear model, allowing us to interpret the coefficients as with a linear dependent model (which yields very similar results). Standard errors are reported as White robust standard errors.
  Dashed horizontal lines separate the different categorical variables.
  Pseudo R$^2$ has been computed as $1- \text{Residual Deviance} / \text{Null Deviance}$.
  The sample is restricted to already-existing firms as most covariates don't make sense for companies without power asset ownership. 
  Each column represents the result from subsetting different time windows. Note that the time horizon for the full sample is 2006 to 2023 since we require the horizon 2001--2005 to compute the \emph{Invested x in past 5 yrs} variable. 
  } 
  \label{t:reg_invest} 
\end{table}

\FloatBarrier

\newpage
\subsection{Drivers of technology choice}

If we know that a firm invests in a given year, can we infer which technology it will invest in?
To answer this question, we fit logistic regression models where the dependent variable is the binary variable indicating whether a given firm adopts technology $k$, conditional on the firm investing in that year.
Again, we include the constant, time effects, and the total power capacity owned.
We further include two key covariates in our model. The first, emph{focal tech share}, represents the relative share of the focal technology $k$ within a firm's portfolio (grouped into buckets). The second is a categorical variable indicating the investment behavior of firms over the past five years: (a) no investment, (b) investment in the focal technology only, (c) investment in both the focal technology and other technologies, and (d) investment exclusively in other technologies.\footnote{
We have also experimented with including the firms' capacity ownership of the focal technology $k$. However, including both total firm capacity and technology $k$ capacity introduces collinearity issues as evidenced by a substantial increase in variance inflation factors. Consequently, we have not included technology $k$ specific capacities but only include total firm capacity. Note that still control for the relevance of technology $k$ in firm's portfolio trough the categorical covariate \emph{focal tech share}.
}

We fit the logistic model for each focal technology $k$ separately and report marginal effects in Table~\ref{t:regall_type}. 
Note that the number of investments observed varies greatly across different technology classes. While about a third of all investments (at the level of firm-technology-year triples) over the past ten years are solar PV capacity additions, we observe only 92 investments in solar thermal and 130 investments in nuclear power.
Depending on the technology, these models can explain between 22\% (solar thermal) and 53\% (nuclear).

\begin{figure}[H]
    \centering
    \includegraphics[width=\linewidth]{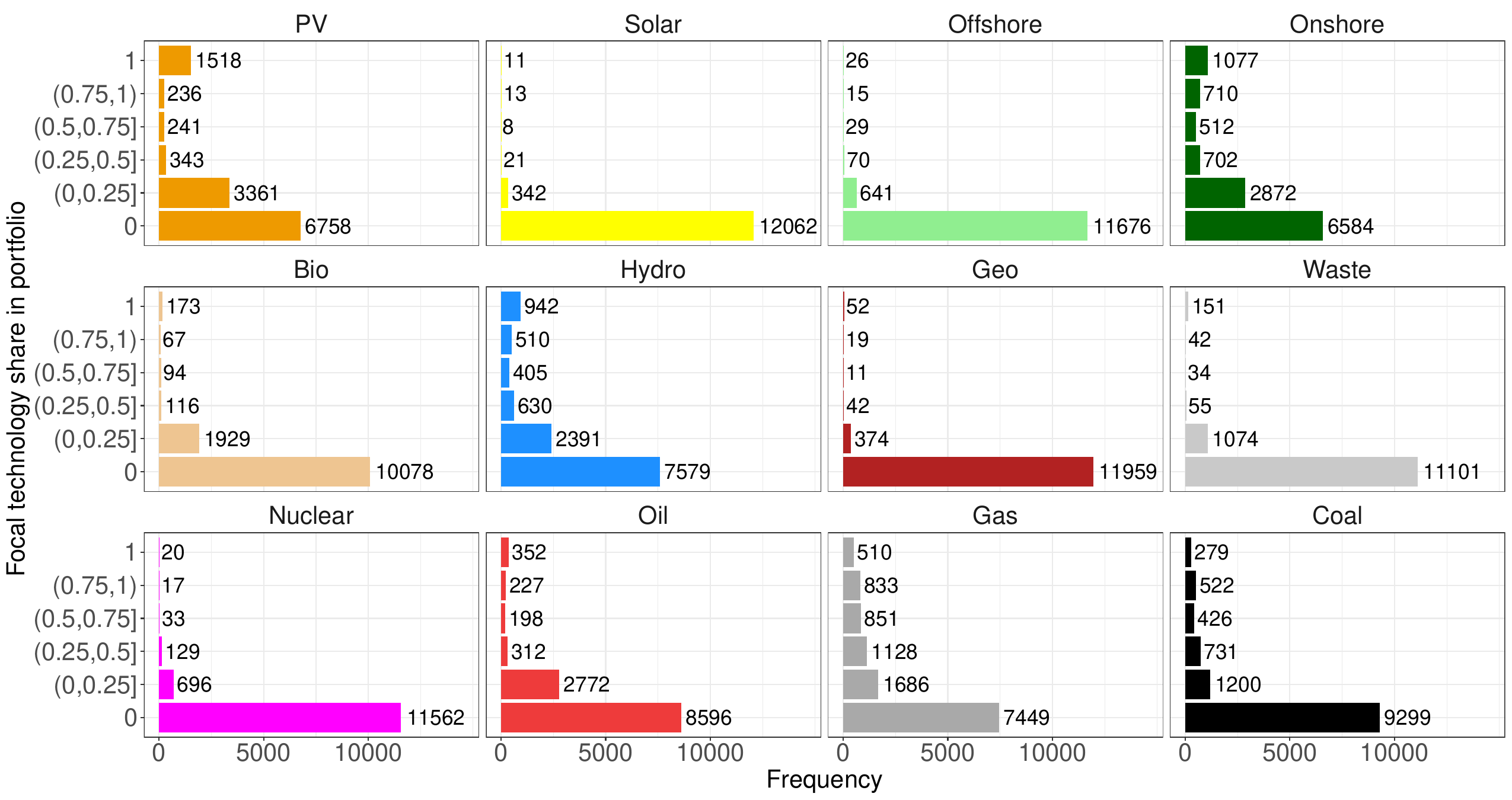}
    \caption{
    {\bf Frequency distribution of technology share buckets}. The figure is based on the sample used in the logistic regression shown in Table~\ref{t:regall_type}.
    Numbers next to the columns indicate the number of observations for each bucket.
    }
    \label{fig:bucket}
\end{figure}

Our results show that a technology's relative share within a firm's portfolio is a key factor in determining whether an investor adds additional capacity of that particular technology. If a firm holds more of a given technology in its portfolio, the more likely it is to invest again in this technology. This holds essentially across all technologies although for some technologies we have limited variance to explore (Fig.~\ref{fig:bucket}), making it difficult to establish significant relationships for all buckets. For example, we have only very few observations of investing firms that hold more than 25\% of solar thermal in their portfolio.

Note that there are also some differences in effect size. For example, holding only geothermal or coal capacities increases the probability of investing in geothermal and coal, respectively, by about 50\%. For offshore wind and hydropower, the effect is substantially larger (about 63\% for both).
In any case, our analysis reveals a strong correlation between an investor's technology portfolio composition and its subsequent investment decisions. Importantly, this relationship is not linear but intensifies with higher portfolio concentration.

The second key explanatory variable is the historical investment behavior of a firm.\footnote{
However, for technological categories with few observations and small number of previous investments it is again difficult to establish significant relationships.
}
Compared to a firm that has not invested at all in the last five years, the likelihood of investing in a particular technology increases substantially if the investor has added capacity of that technology already in the preceding years.
The likelihood of reinvesting in a particular technology is the highest if the firm has previously invested exclusively in that technology. This is particularly true for firms that have only invested in nuclear in the past five years, as indicated by an extremely large coefficient of 0.99. Although much smaller than for nuclear, we also observe large marginal effect sizes for solar PV and wind onshore investments.   
Analogously, if the investor has previously invested exclusively in other technologies, the probability of investing in the focal technology tends to decline.

We find mixed effects of investor types (private, public or non-firm entities) on the adoption of specific technologies. For example, conditional on investing, non-firm entities are more likely to adopt solar pv or biopower than private or public firms. In contrast, public firms are more likely to adopt onshore wind capacities than other investor types and private firms are more likely to invest in oil. For the adoption of gas, waste and hydro assets, the firm type does not have any significant impact. 

We further run simple linear regression models to investigate the relationship between firm size (in total capacity owned) and the investment size in focal technology $k$, conditional on investing in this technology. As shown in Table~\ref{t:regall_size}, investment amounts are closely related to the total capacity ownership of firms. The larger a firm, the more it will invest in a given technology (if it invests at all in this technology). 
For example, consider a firm investing in solar PV. A 1\% increase in total capacity of this firm, is associated with a 0.53\% increase in solar pv investment. The size-investment relationship is remarkably similar across different technological classes. 

We have also experimented with including further covariates, such as shown in the other regression tables. We found that including other variables can improve the model fit somewhat. However, total firm log capacity is by far the most important covariate, explaining up to 64\% of the observed variance. Thus, we have decided to show the simplest model specification focusing on the key covariate.

\begin{landscape}

\begin{table}[!htbp] \centering 
\footnotesize
\begin{tabular}{@{\extracolsep{2pt}}lcccccccccccc} 
\\[-1.8ex]\hline 
\hline \\[-1.8ex] 
\textit{Dependent variable:}& \multicolumn{12}{c}{Firm-specific likelihood to invest in technology $k$ in a given year, conditional on investing}  \\ 
Focal technology $k$ & Solar & Solar & Offshore & Onshore & Bio & Hydro & Geo & Waste & Nuclear & Oil & Gas & Coal \\ 
& PV & thermal & wind & wind & & & & & & & & \\ 

\hline \\[-1.8ex] 
Constant \& year dummies   &included &included &included &included  &included &included &included &included &included  &included &included &included \\[1.2ex] 
 Log total capacity & 0.011$^{**}$ & 0.001 & 0.004$^{**}$ & 0.005$^{**}$ & $-$0.0005 & 0.006$^{**}$ & 0.0004 & 0.001 & 0.002$^{**}$ & $-$0.0001 & 0.012$^{***}$ & 0.007$^{**}$ \\ 
  & (0.002) & (0.0003) & (0.001) & (0.002) & (0.001) & (0.001) & (0.0004) & (0.001) & (0.001) & (0.001) & (0.001) & (0.001) \\ 
  \hdashline
  Baseline: Private firm \\[1.2ex]
  Public firm & $-$0.003 & $-$0.004 & $-$0.001 & 0.036$^{**}$ & 0.015 & 0.023$^{*}$ & 0.002 & 0.002 & 0.002 & $-$0.013$^{*}$ & $-$0.010 & 0.022$^{*}$ \\ 
  & (0.014) & (0.003) & (0.006) & (0.013) & (0.009) & (0.009) & (0.003) & (0.005) & (0.003) & (0.006) & (0.011) & (0.010) \\ 
  Other entity & 0.052$^{**}$ & $-$0.002 & 0.011 & 0.021 & 0.037$^{**}$ & 0.0003 & 0.005 & 0.006 & $-$0.005 & $-$0.025$^{**}$ & $-$0.003 & 0.023$^{*}$ \\ 
  & (0.015) & (0.003) & (0.007) & (0.014) & (0.012) & (0.010) & (0.004) & (0.006) & (0.003) & (0.006) & (0.011) & (0.010) \\ 
      \hdashline
Baseline: No investments \\
(past 5 yrs) \\[1.2ex]
  Invested in focal \& other tech  & 0.119$^{**}$ & 0.005 & 0.004 & 0.121$^{**}$ & 0.014 & 0.023$^{*}$ & 0.010 & 0.020$^{**}$ & 0.004 & 0.027$^{**}$ & $-$0.017 & 0.008 \\ 
  (past 5 yrs) & (0.020) & (0.007) & (0.007) & (0.016) & (0.007) & (0.009) & (0.006) & (0.008) & (0.005) & (0.007) & (0.010) & (0.007) \\ 
  Invested in focal tech only & 0.236$^{**}$ & $-$0.0005 & 0.023 & 0.163$^{***}$ & 0.039$^{*}$ & 0.110$^{**}$ & 0.029 & 0.097$^{**}$ & 0.990$^{***}$ & 0.051$^{**}$ & 0.083$^{**}$ & 0.041$^{**}$ \\ 
  (past 5 yrs)& (0.029) & (0.007) & (0.026) & (0.018) & (0.015) & (0.015) & (0.017) & (0.028) & (0.001) & (0.012) & (0.015) & (0.012) \\ 
  Invested in other tech only & $-$0.073$^{**}$ & 0.003 & $-$0.0005 & $-$0.071$^{**}$ & $-$0.025$^{**}$ & $-$0.066$^{**}$ & $-$0.004 & $-$0.011$^{*}$ & $-$0.007 & $-$0.053$^{**}$ & $-$0.119$^{***}$ & $-$0.056$^{**}$ \\ 
  (past 5 yrs) & (0.012) & (0.003) & (0.005) & (0.013) & (0.006) & (0.010) & (0.004) & (0.004) & (0.005) & (0.007) & (0.010) & (0.008) \\ 
      \hdashline
Baseline: Focal tech share: 0 \\[1.2ex]
  Focal tech share: (0,0.25] & 0.096$^{**}$ & 0.053$^{**}$ & 0.119$^{**}$ & 0.203$^{***}$ & 0.070$^{**}$ & 0.130$^{***}$ & 0.090$^{**}$ & 0.041$^{**}$ & 0.036$^{**}$ & 0.076$^{***}$ & 0.131$^{***}$ & 0.078$^{**}$ \\ 
  & (0.015) & (0.020) & (0.021) & (0.013) & (0.009) & (0.008) & (0.022) & (0.008) & (0.009) & (0.007) & (0.011) & (0.010) \\ 
  Focal tech share: (0.25,0.5] & 0.227$^{***}$ & 0.119 & 0.168$^{**}$ & 0.278$^{***}$ & 0.263$^{**}$ & 0.265$^{***}$ & 0.317$^{**}$ & 0.113$^{**}$ & 0.074$^{**}$ & 0.177$^{**}$ & 0.224$^{***}$ & 0.160$^{***}$ \\ 
  & (0.025) & (0.078) & (0.044) & (0.018) & (0.039) & (0.017) & (0.081) & (0.035) & (0.020) & (0.021) & (0.015) & (0.015) \\ 
  Focal tech share: (0.5,0.75] & 0.339$^{***}$ & 0.204 & 0.175$^{*}$ & 0.305$^{***}$ & 0.288$^{**}$ & 0.314$^{***}$ & 0.352$^{*}$ & 0.142$^{**}$ & 0.116$^{*}$ & 0.287$^{***}$ & 0.245$^{***}$ & 0.182$^{***}$ \\ 
  & (0.032) & (0.164) & (0.069) & (0.020) & (0.042) & (0.020) & (0.142) & (0.051) & (0.047) & (0.028) & (0.016) & (0.018) \\ 
  Focal tech share: (0.75,1) & 0.391$^{***}$ & 0.693$^{**}$ & 0.375$^{**}$ & 0.427$^{***}$ & 0.417$^{**}$ & 0.392$^{***}$ & 0.647$^{**}$ & 0.663$^{**}$ & 0.209$^{*}$ & 0.280$^{***}$ & 0.346$^{***}$ & 0.306$^{***}$ \\ 
  & (0.033) & (0.121) & (0.111) & (0.020) & (0.059) & (0.020) & (0.131) & (0.116) & (0.085) & (0.027) & (0.017) & (0.022) \\ 
  Focal tech share: 1 & 0.550$^{***}$ & 0.328 & 0.625$^{**}$ & 0.509$^{***}$ & 0.562$^{***}$ & 0.627$^{***}$ & 0.499$^{**}$ & 0.465$^{**}$ & 0.456 & 0.353$^{***}$ & 0.467$^{***}$ & 0.500$^{***}$ \\ 
  & (0.023) & (0.179) & (0.138) & (0.018) & (0.051) & (0.022) & (0.132) & (0.070) & (0.266) & (0.030) & (0.023) & (0.038) \\ 
 \hline \\[-1.8ex] 
Pseudo R$^2$ & 0.347 & 0.226 & 0.293 & 0.372 & 0.301 & 0.475 & 0.509 & 0.48 & 0.531 & 0.353 & 0.344 & 0.439 \\ 
Nr of investments in $k$ & 4,334 & 92 & 333 & 4,069 & 659 & 2,124 & 202 & 368 & 130 & 800 & 2,116 & 1,077 \\ 
Observations (nr of investments) & 12,457 & 12,457 & 12,457 & 12,457 & 12,457 & 12,457 & 12,457 & 12,457 & 12,457 & 12,457 & 12,457 & 12,457 \\ 
\hline 
\hline \\[-1.8ex] 
\textit{Note:}  & \multicolumn{12}{r}{$^{*}$p$<$0.05; $^{**}$p$<$0.01; $^{***}$p$<$1e-16} \\ 
\end{tabular} 
  \caption{
  \textbf{Marginal effects from a logistic regression of the binary dependent variable (investing in a focal technology: Yes/No), onto a set of covariates.} 
  As in Table~\ref{t:reg_invest}, we report average marginal effects, allowing us to interpret the coefficients as with a linear dependent model. Standard errors are reported as White robust standard errors.
  Dashed horizontal lines separate the different categorical variables.
  Pseudo R$^2$ has been computed as $1- \text{Residual Deviance} / \text{Null Deviance}$.
  Each column represents the result when fixing a different focal technology $k$. 
  The sample is restricted to already-existing firms as most covariates don't make sense for companies without power asset ownership. Results are based on observations of the most recent decade in our sample. This results in a total of 12.5K investment observations. \emph{Nr of investments in $k$} indicates the number of investments observed per focal technology $k$.
  } 
  \label{t:regall_type} 
\end{table} 
\end{landscape}

\begin{landscape}
\begin{table}[!htbp] \centering 
\footnotesize
\begin{tabular}{@{\extracolsep{2pt}}lcccccccccccc} 
\\[-1.8ex]\hline 
\hline \\[-1.8ex] 
\textit{Dependent variable:} & \multicolumn{12}{c}{Log capacity investment in tech $k$, conditional on investing in tech $k$} \\ 
Focal technology $k$ & PV & Solar & Offshore & Onshore & Bio & Hydro & Geo & Waste & Nuclear & Oil & Gas & Coal \\ 
\hline \\[-1.8ex] 
Year dummies   &included &included &included &included  &included &included &included &included &included  &included &included &included \\[1.2ex] 
 Log total capacity & 0.526$^{***}$ & 0.605$^{***}$ & 0.529$^{***}$ & 0.614$^{***}$ & 0.565$^{***}$ & 0.615$^{***}$ & 0.663$^{***}$ & 0.584$^{***}$ & 0.596$^{***}$ & 0.662$^{***}$ & 0.613$^{***}$ & 0.551$^{***}$ \\ 
  & (0.009) & (0.073) & (0.038) & (0.011) & (0.020) & (0.012) & (0.055) & (0.032) & (0.038) & (0.019) & (0.013) & (0.022) \\ 
  Constant & 0.473$^{**}$ & 1.245 & 1.063 & $-$0.188$^{*}$ & 0.180 & $-$0.190 & 0.064 & 0.278 & 1.801$^{**}$ & $-$0.489$^{**}$ & 0.737$^{**}$ & 1.560$^{**}$ \\ 
  & (0.103) & (0.906) & (0.649) & (0.094) & (0.160) & (0.111) & (0.550) & (0.260) & (0.344) & (0.189) & (0.142) & (0.208) \\ 
 \hline \\[-1.8ex] 
Observations (nr of investments in $k$) & 4,334 & 92 & 333 & 4,069 & 659 & 2,124 & 202 & 368 & 130 & 800 & 2,116 & 1,077 \\ 
Adjusted R$^{2}$ & 0.454 & 0.470 & 0.457 & 0.509 & 0.515 & 0.556 & 0.480 & 0.484 & 0.639 & 0.580 & 0.566 & 0.530 \\ 
\hline 
\hline \\[-1.8ex] 
\textit{Note:}  & \multicolumn{12}{r}{$^{*}$p$<$0.05; $^{**}$p$<$0.01; $^{***}$p$<$1e-16} \\ 
\end{tabular} 
  \caption{
    \textbf{Linear regression estimates for technology-specific investment size, conditional on investing in this technology.} 
    Standard errors are reported as White robust standard errors.
  Each column represents the result when fixing a different focal technology $k$. 
  The sample is restricted to firms that invest in a given year in the focal technology (column). Results are based on observations of the most recent decade in our sample.
  } 
  \label{t:regall_size} 
\end{table} 
\end{landscape}

\newpage
\FloatBarrier
\small
\bibliographystyle{agsm}
\bibliography{../references}